\documentclass[12pt]{article}
\usepackage{amsfonts}
\usepackage{latexsym}
\usepackage {ulem}
\usepackage{amsmath,amssymb,braket}
\usepackage{verbatim}
\usepackage{cite}
\usepackage[colorlinks=true
            	,urlcolor=blue
            	,anchorcolor=blue
            	,citecolor=blue
            	,filecolor=blue
            	,linkcolor=blue
            	,menucolor=blue
            	,linktoc=page
            	]{hyperref}

\usepackage[T1]{fontenc}
\usepackage[textsize=footnotesize,textwidth=2.5cm]{todonotes}
\newcommand{\wei}[2]{\textcolor{magenta}{#1}\todo[color=yellow]{\scriptsize{WS: #2}}}

\newcommand{\hs}[2]{\textcolor{magenta}{#1}\todo[color=yellow]{\scriptsize{HS: #2}}}

\usepackage{geometry}

\newcommand{\bea}{\begin{eqnarray}}
\newcommand{\eea}{\end{eqnarray}}
\newcommand{\be}{\begin{equation}}
\newcommand{\ee}{\end{equation}}
\newcommand{\ba}{\begin{array}}
\newcommand{\ea}{\end{array}}
\newcommand{\baa}{\begin{aligned}}
\newcommand{\eaa}{\end{aligned}}

\newcommand{\Red}[1]{\textcolor{red}{#1}}

\def\hl{h_{\tilde \lambda}}
\def\bhl{{\bar h}_{\tilde \lambda }}
\def\nn{\nonumber}

\def\p{\partial}

\newcommand{\eq}[1]{\begin{align}#1\end{align}}

\numberwithin{equation}{section}
\begin{document}
\begin{titlepage}
\begin{center}
{\Large \bf Correlation Functions in the TsT/$T{\bar T}$ Correspondence }
\lineskip .75em
 \vskip 2.5cm
  { 
  Wei Cui$^{a,b,}$\footnote{cweiphys@gmail.com}, Hongfei Shu$^{a,b,}$\footnote{shuphy124@gmail.com}, Wei Song$^{b,}$\footnote{wsong2014@mail.tsinghua.edu.cn}
  and Juntao Wang$^{a,b,}$\footnote{jtwang.bimsa@gmail.com} }
\vskip 2.5em
 {
 \it 
$^a$Yanqi Lake 
Beijing Institute of Mathematical Sciences and Applications (BIMSA), 
Huairou District, 
Beijing 101408, 
China\\
$^b$Yau Mathematical Sciences Center, Tsinghua University, Beijing 100084, China
}
\vskip 3.0em
\end{center}
\begin{abstract}

We investigate the proposed holographic duality between the TsT transformation of IIB string theory on AdS$_3\times {\cal N}$ with NS-NS flux and a single-trace $T\bar{T}$ deformation of the symmetric orbifold CFT. We present a non-perturbative calculation of two-point correlation functions using string theory and demonstrate their consistency with those of the $T\bar{T}$ deformation. The two-point correlation function of the deformed theory on the plane, written in momentum space, is obtained from that of the undeformed theory by replacing $h$ with $h+2{\tilde \lambda\over w} p\bar p$, where $h$ is the spacetime conformal weight, $\tilde \lambda$ is a deformation parameter, $p$ and $\bar p$ are the momenta, and $w$ labels the twisted sectors in the deformed symmetric product. At $w=1$, the non-perturbative result satisfies the Callan-Symanzik equation for double-trace $T\bar T$ deformed CFT derived in \cite{Cardy:2019qao}. We also perform conformal perturbations on both the worldsheet CFT and the symmetric orbifold CFT as a sanity check. The perturbative and non-perturbative matching between results on the two sides provides further evidence of the conjectured TsT/$T\bar{T}$ correspondence.

  \end{abstract}
\end{titlepage}

\newpage

\setcounter{tocdepth}{3}

\tableofcontents
\bigskip

\section{Introduction}

The AdS$_{d+1}$/CFT$_d$ correspondence relating gravity theory in ($d+1$)-dimensional Anti-de Sitter (AdS) spacetime to a conformal field theory (CFT) on its boundary is one of the most profound results in string theory~\cite{Maldacena:1997re}. 
Through a holographic dictionary, observables of gravity theory in AdS$_{d+1}$ are mapped to those of the CFT$_d$~\cite{Gubser:1998bc, Witten:1998qj}. In particular, the Bekenstein-Hawking entropy of some special black holes can be accounted for from counting the microstates in CFTs ~\cite{Strominger:1996sh, Strominger:1997eq}. 
While the AdS/CFT correspondence has solved many conceptual problems in a highly illuminating and efficient way, it is not directly applicable to most physics in the real world where non-AdS spacetimes and non-conformal field theories are ubiquitous.

One way to extend the holographic principle beyond the AdS/CFT correspondence is to deform both sides of some known examples of holographic duality constructed in string theory. The advantage of this top-down approach is that string theory is a UV-complete theory and hence provides more tools for a systematic study.  
Consider IIB string theory on AdS$_3\times {\cal N}$ with pure NS-NS flux, which can be obtained as the near horizon limit of $k$ parallel NS-5 branes and $N$ fundamental strings. 
String theory on this background has a weakly coupled worldsheet description in terms of the $SL(2,\mathbb{R})_k$ Wess-Zumino-Witten (WZW) model~\cite{Maldacena:2000hw,Maldacena:2000kv,Maldacena:2001km},
the spectrum of which includes long string sectors as well as short string sectors. 
It has been conjectured
\cite{Argurio:2000tb,
Giribet:2018ada,Gaberdiel:2018rqv,Eberhardt:2019ywk,Eberhardt:2019qcl,Eberhardt:2021vsx} that the long string sector of string theory on the aforementioned AdS$_3\times {\cal N}$ background is holographically dual to a symmetric orbifold CFT ${\rm Sym}^N(\mathcal{M}_0)$ where $\mathcal{M}_0$ is a seed CFT with central charge $6k$ and $N$ is the number of fundamental strings in the background.

The conjectured AdS$_3$/CFT$_2$ duality has undergone various checks including matching the spectrum and correlation functions~\cite{Giribet:2018ada,Gaberdiel:2018rqv,Eberhardt:2018ouy,Eberhardt:2021vsx,Eberhardt:2019qcl, Eberhardt:2019ywk}, and thus provides a good starting point to study deformations.  
In this paper, we focus on the single-trace $T\bar T$ deformation \cite{Giveon:2017nie} of the symmetric orbifold CFT, and correspondingly a T dual-shift-T dual (TsT) transformation \cite{Apolo:2018qpq,Araujo:2018rho, Apolo:2019zai} in string theory \footnote{The relation between TsT transformation and single trace $T\bar J$ deformation was observed in \cite{Apolo:2018qpq}. Single trace $T\bar T$ and $T\bar J$ deformation were interpreted as $O(d,d)$ transformations in \cite{Araujo:2018rho}. A systematic study of the TsT/$T\bar T$ correspondence and its generalization to $T{\bar T}+J{\bar T}+T{\bar J}$ can be found in \cite{Apolo:2019zai,Apolo:2021wcn}}. 
In the following, we will describe the two deformations in more details.

To define a single-trace $T\bar T$ deformation, we need to first introduce the double-trace $T\bar T$ deformation\footnote{In the literature, $T\bar T$ deformation often refers to the double-trace version. }~\cite{Zamolodchikov:2004ce,Smirnov:2016lqw, Cavaglia:2016oda} which is an irrelevant but solvable deformation in the sense that many of the physical quantities including the spectrum and the scattering amplitudes can be computed in terms of those before the deformation\cite{Smirnov:2016lqw, Cavaglia:2016oda,Cardy:2018sdv,Bonelli:2018kik,Dubovsky:2018bmo,Conti:2018tca,Coleman:2019dvf,Hashimoto:2019wct}. The deformed theory has several remarkable features including factorization, intriguing reformaluations as flat Jackiw-Teitelboim gravity \cite{Dubovsky:2017cnj,Dubovsky:2018bmo}, random metrics \cite{Cardy:2018sdv,Hirano:2020nwq} and string theory \cite{Callebaut:2019omt,Tolley:2019nmm}, and modularity and universality of torus partition functions \cite{Cardy:2018sdv,Datta:2018thy, Aharony:2018bad, Apolo:2023aho}.
Progresses have also been made in correlation functions \cite{Kraus:2018xrn,Cardy:2019qao,He:2019vzf,He:2020udl,Hirano:2020nwq,Hirano:2020ppu,He:2020qcs,He:2022jyt} and entanglement entropy \cite{Donnelly:2018bef,Chen:2018eqk,Sun:2019ijq,Jeong:2019ylz,He:2022xkh}.  
The holographic dual has been proposed to be Einstein gravity on AdS$_3$ after cuting-off the asymptotic region \cite{McGough:2016lol,Kraus:2018xrn} or gluing a patch of an auxillary locally AdS$_3$ spacetime to the original one \cite{Apolo:2023vnm}, depending on the sign of the deformation parameter. An alternative proposal is to impose mixed boundary conditions at the asymptotic boundary of AdS$_3$ spacetime \cite{Bzowski:2018pcy, Guica:2019nzm}. See also \cite{Kawamoto:2023wzj, Jafari:2019qns} for related proposals and also further research \cite{Khoeini-Moghaddam:2020ymm} motivated by this correspondence.

The double-trace $T\bar T$ deformation is universal and can be defined for any quantum field theory that has two translational symmetries so that the stress tensor can be defined. In particular, we can apply the  
double-trace $T\bar T$ deformation to the seed CFT $\mathcal{M}_0$ of the symmetric orbifold CFT. 
The single-trace $T\bar T$ deformed  symmetric orbifold denoted by ${\rm Sym}^N(\mathcal{M}_\mu)$ 
can then be constructed from the deformed seed theory $\mathcal{M}_\mu$ by the usual orbifolding procedure. More explicitly, the deformed action at deformation parameter $\mu$ satisfies the following differential equation \footnote{In contrast to the single-trace version \eqref{singlettbar},  the double-trace $T\bar T$ deformation of the symmetric orbifold CFT is defined as, \be\frac{\partial S_{\mu}}{\partial \mu}=-\sum_{I,I'=1}^N\frac{1}{\pi}\int  J^I_{(1)}\wedge J^{I'}_{(\bar{2})}.\ee } \begin{equation}\label{singlettbar}
 \frac{\partial S_{\mu}}{\partial \mu}=-\sum_{I=1}^N\frac{1}{\pi}\int  J^I_{(1)}\wedge J^I_{(\bar{2})},
 \end{equation}
where 
\begin{equation}\label{JJbar}
J^I_{(1)}=T^I_{xx}dx+T^I_{\bar{x}x}d\bar{x},\quad J^I_{(\bar{2})}=T^I_{x\bar{x}}dx+T^I_{\bar{x}\bar{x}}d\bar{x}
\end{equation}
are the Noether currents generating translational symmetries on the $I$-th copy of the seed CFT $\mathcal{M}_\mu$. 
The spectrum in the untwisted sector of the single-trace $T\bar T$ deformed CFT takes the same form as that in the double-trace version \cite{Giveon:2017nie}, the spectrum in the twisted sector has been worked out assuming modular invariance of the torus partition function \cite{Apolo:2023aho}, and a UV completion exhibiting strong-weak coupling duality has been proposed in \cite{Benjamin:2023nts}. Although the single-trace $T\bar T$ deformation is only explicitly defined for symmetric orbifold CFTs, it is plausible to expect a similar deformation for more general CFTs. See \cite{Manschot:2022lib} for a related discussion for the MSW CFT \cite{Maldacena:1997de}.

The single-trace $T\bar{T}$-deformed CFT ${\rm Sym}^N(\mathcal{M}_\mu)$ \eqref{singlettbar} is conjectured to be holographically dual to  the long string sector of IIB string theory on a deformed background
generated by TsT transformation \cite{Apolo:2019zai} along the two lightcone directions of AdS$_3$.
Unlike the double-trace case, the bulk metric in the single-trace case is no longer locally AdS$_3$ but interpolates between the linear dilaton background at the UV and AdS$_3$ at the IR~\cite{Giveon:2017nie}, and thus the TsT/$T\bar T$ correspondence provides an explicit example of holographic duality beyond AdS. 
It has been shown that the string spectrum calculated from the string worldsheet theory \cite{Giveon:2017nie,Apolo:2019zai} agrees with the spectrum of ${\rm Sym}^N(\mathcal{M}_\mu)$ both in the untwisted sector and twisted sectors \cite{Apolo:2023aho}, the torus partition function from string theory \cite{Hashimoto:2019hqo} agrees with that derived  from the symmetric product \cite{Apolo:2023aho} and both can be written in terms of generalized Hecke operators, and the black hole entropy and thermodynamics can also be reproduced from the dual field theory analysis \cite{Apolo:2019zai}.

In this paper, we will  provide  further evidence for the proposed TsT/$T\bar T$ duality by  showing that the correlation functions calculated from both sides indeed match each other. 
The calculation of correlation functions of $T\bar T$ deformed CFT has been a challenge, even for the more well-studied double-trace version. As the deformation is irrelevant and introduces non-locality, it is difficult to write a compact form in position space. Nevertheless, 
a flow equation and a Callan-Symanzik equation have been written down \cite{Cardy:2019qao}, the later of which can be conveniently solved in momentum space with appropriate boundary conditions.  

One of the main results of this paper is that we provide a non-perturbative result of two-point functions from the string worldsheet theory after the TsT transformation. 
The basic idea is as follows: by a field redefinition \cite{Alday:2005ww} strings on the TsT transformed background can be mapped to strings on AdS$_3$ background with twisted boundary conditions, the latter of which can be further cast into a momentum-dependent spectra flow transformation \cite{Azeyanagi:2012zd,Apolo:2019zai}. In this paper we implement the momentum-dependent spectra flow transformation to vertex operators. 
Consider a vertex operator $\tilde{V}^w_{j,h} (z, p)$ in AdS$_3$ background with spacetime conformal weights $(h,\bar h)$, winding number $w$ and momentum $(p,\bar p)$, the deformed operator denoted by $\hat{V}^w_{j,h} (z, p)$ will get an extra dressing factor, with a modified weight $h_{\tilde \lambda }$ determined by
\begin{equation}\label{hfunctionintro}
h\xrightarrow[]{\text{TsT}}h_{\tilde \lambda }= h +2{\tilde{\lambda} \over {w} } p \bar p
\end{equation}
where $\tilde{\lambda}$ is the TsT parameter which is to be identified with the $T\bar T$  parameter $\mu$ on some units.
Then it can be shown explicitly that after TsT the only change in the momentum space correlation function is to replace the weight from constant $h$ to $h_{\tilde \lambda }$, that is:
\begin{equation}\label{2ptchange}
\left \langle\tilde{V}^w_{j,h} (z, p) \tilde{V}^w_{j,h} (z, -p)\right \rangle\sim (p\bar{p})^{2h-1}\xrightarrow[]{\text{TsT}}\left \langle\hat{V}^w_{j,h} (z, p) \hat{V}^w_{j,h} (z, -p)\right \rangle\sim (p\bar{p})^{2h-1+4{\tilde{\lambda} \over {w} }p\bar{p}},
\end{equation}
where we have omitted some normalization factors. As a consistency check, we expand the non-perturbative result \eqref{2ptchange} at small $\tilde{\lambda}$, and find that it indeed agrees with the perturbative calculation on the string worldsheet. 

Using the holographic dictionary, a vertex operator with winding number $w$ corresponds to an operator in the $w-$twisted sector of the symmetric product theory. In particular, $w=1$ corresponds to the untwisted sector in which results for the single-trace and double-trace deformations are expected to be the same. Indeed, our non-perturbative two-point function \eqref{2ptchange} with $w=1$ does take a similar form as that in the double-trace $T\bar T$ deformation, and we can check explicitly that it satisfies the Callan-Symanzik equation \cite{Cardy:2019qao}.
For arbitrary $w$, we perform a perturbative computation for the two-point function of the twisted operators in the deformed orbifold theory ${\rm Sym}^N(\mathcal{M}_\mu)$, and find that the the result coincides with the first-order result obtained from string theory. 
Our results thus provides a further test of the conjected TsT/single-trace $T\bar{T}$ correspondence. 

The paper is organized as follows. In section 2, we review the relation between TsT transformation and $T\bar T$ deformation. In section 3, we employ the the spectral flow techniques to derive the momentum dependent weight for vertex operators after the TsT transformation.  Two-point correlation functions from the worldsheet and deformed symmetric orbifold are derived in Section 4 and section 5, respectively. 
In section 6, we draw a conclusion and discuss possible future directions.

\section{TsT and Single-Trace $T\bar T$  Deformation}

In this section, we will briefly review the TsT transformation in string theory and the single-trace $T\bar{T}$ deformation, and set up the notation in this paper.

Let us begin with the bosonic string theory with the target space metric $\tilde G_{\mu\nu}$ and a Kalb-Ramond background field $\tilde B_{\mu\nu}$:
\begin{equation}\label{action}
\tilde{S}=-\frac{1}{4\pi l^2_s}\int d^2z(\sqrt{-\eta}\eta^{ab}\tilde{G}_{\mu\nu}+\epsilon^{ab}\tilde{B}_{\mu\nu})\partial_a\tilde{X}^{\mu}\partial_b\tilde{X}^{\nu},
\end{equation}
where the worldsheet coordinates are $z=\tau+\sigma, \bar{z}=\tau-\sigma$ and 
\begin{equation}
\eta=\textnormal{det}(\eta_{ab}),\quad \eta_{z\bar{z}}=-\frac{1}{2},\quad \eta^{z\bar{z}}=-2, \quad \epsilon_{z\bar{z}}=-\epsilon^{z\bar{z}}=1.
\end{equation}
The action can be further written more compactly as:
\begin{equation}\label{action1}
\tilde{S}=\frac{1}{2\pi l^2_s}\int d^2z\tilde{M}_{\mu\nu}\partial\tilde{X}^{\mu}\bar{\partial}\tilde{X}^{\nu},
\end{equation}
where 
$\tilde{M}_{\mu\nu}=\tilde{G}_{\mu\nu}+\tilde{B}_{\mu\nu}$.
Let us assume that the background has two  explicit $U(1)$ isometries which generate translations along the directions parameterized by $\tilde{X}^1$ and $\tilde{X}^{\bar{2}}$. The two $U(1)$ symmetries are global symmetries on the string worldsheet, whose corresponding Noether currents can be schematically written as:
\begin{equation}\label{noecurrent}
\tilde{j}^a_{(n)}=-l^{-2}_s(\sqrt{-\eta}\eta^{ab}\tilde{G}_{n\mu}\partial_b\tilde{X}^{\mu}+\epsilon^{ab}\tilde{B}_{n\mu}\partial_b\tilde{X}^{\mu}),\quad n=1,\bar{2}
\end{equation}
or in terms of differential form,
\begin{equation}
 \tilde{{\bf{j}}}_{(n)}\equiv 2 \tilde{j}_{(n)}^a \eta_{ab} d z^b =  \big(\tilde{G}_{n\mu} d \tilde{X}^{\mu}  - \tilde{B}_{n\mu } \star d \tilde{X}^{\mu}\big).
\end{equation}
Now we perform a TsT transformation by first doing a T-duality along $\tilde X^1$, followed by a shift $\tilde X^{\bar 2} = X^{\bar 2}-2\tilde \lambda \tilde X^1 $ and finally another T-duality along $\tilde X^1$.
One key point established in \cite{Apolo:2019zai} is that the TsT transformation is equivalent to the deformation of worldsheet action by the anti-symmetric product of two Noether currents:
\begin{equation}{\label{jjdeform}}
    \frac{\partial S}{\partial \tilde{\lambda}}={-\frac{1}{\pi}}\int {\bf{j}}_{(1)}\wedge{\bf{j}}_{(\bar{2})}
\end{equation}
where ${\bf{j}}_{(n)}$ is the Noether current after the TsT transformation. As we will see momentarily, formula (\ref{jjdeform}) will be very illuminating in the comparison between TsT and $T\bar{T}$ deformation. 

It has been conjectured that the long string sector of IIB string theory on AdS$_3\times {\cal N}$ with NS-NS flux whose worldsheet theory in the form of \eqref{action}
is holographically dual to a symmetric product CFT ${\rm Sym}^N(\mathcal{M}_0)$, where ${\cal M}_0$ is a seed CFT with central charge $6k$. 
The single-trace $T\bar T$ deformed theory in this context retains the structure of symmetric orbifold, but with a new seed $\mathcal{M}_\mu$ which is the double-trace $T\bar T$ deformation of the original seed $\mathcal{M}_0$.  
As a result, the deformed action of ${\rm Sym}^N(\mathcal{M}_\mu)$ satisfies the following differential equation 
\begin{equation}\label{singlettdeform}
 \frac{\partial S_{\mu}}{\partial \mu}=-\sum_{I=1}^N{\frac{1}{\pi}}\int  J^I_{(1)}\wedge J^I_{(\bar{2})},
 \end{equation}
where $J^I$ is the Noether current degenerating translational symmetry along the two light-cone coordinates $x=x^0+x^1$ and $\bar{x}=x^0-x^1$, 
\begin{equation}
J^I_{(1)}=T^I_{xx}dx+T^I_{\bar{x}x}d\bar{x},\quad J^I_{(\bar{2})}=T^I_{x\bar{x}}dx+T^I_{\bar{x}\bar{x}}d\bar{x}.
\end{equation}
Here $T^I$ is the energy-momentum tensor on the $I$-th copy.
Spectrum in the untwisted sector is similar to the case of double-trace $T\bar T$ deformation \cite{Giveon:2017nie}, and can be written in terms of the undeformed spectrum and deformation parameter. 
Assuming that partition functions on the torus are modular invariant, one can derive the deformed spectrum in the twisted sectors \cite{Apolo:2023aho}.

Holographically, it has been conjectured that (\ref{singlettdeform}) corresponds to TsT deformation \eqref{jjdeform} on the string theory side. Indeed, the similarity between the differential equations  satisfied by the TsT transformation (\ref{jjdeform}) and the single-trace $T\bar T 
 $(\ref{singlettdeform}) suggests that the TsT parameter $\tilde \lambda$ and the $T\bar T$ parameter $\mu$ are related by 
\be
\ell^2\tilde \lambda=\mu\label{dic}
\ee
where $\ell=\sqrt{k}l_s$ is the radius of AdS$_3$.
As evidence, the long string spectrum calculated from the worldsheet theory \cite{Giveon:2017myj,Apolo:2019zai} has been shown to match with that in ${\rm Sym}^N(\mathcal{M}_\mu)$\cite{Apolo:2023aho}, the entropy and thermodynamics of black holes after the TsT transformation can be reproduced by the dual symmetric product theory \cite{Apolo:2019zai}, and more recently torus partition function is also shown to have a universal form at large $N$.  
In the following of this paper, we will study the two-point function of the twisted operators on the both sides, and provide more evidence for this correspondence.

\section{Momentum Dependent Weight From TsT}

In this section, we will review how TsT transformation can induce a momentum dependent spectral flow and as a result, how the weight $h$ will change as a function of deformation parameter $\tilde{\lambda}$ and momentum $p$.

\subsection{String spectrum on AdS$_3\times {\cal N}$}

Let us consider the AdS$_3$ background in Poincar\'{e} coordinates with the metric and $\tilde{B}$ field 
\begin{equation}\label{metricandbform}
d\tilde{s}^2=k(d\rho^2+e^{2\rho}d\tilde{\gamma} d\tilde{\bar{\gamma}}), \quad \tilde{B}=-\frac{k}{2}e^{2\rho} d\tilde{\gamma}\wedge d\tilde{\bar{\gamma}}.
\end{equation} 
The worldsheet action is given by
\begin{equation}\label{oaction}
\tilde{S}=\frac{k}{2\pi}\int d^2 z(e^{2\rho}\partial \tilde{\bar{\gamma}}\bar{\partial}\tilde{\gamma}+\partial\rho\bar{\partial}\rho).
\end{equation}
By introducing an auxiliary field $\beta$, the action can be rewritten as:
\begin{equation}\label{aaction}
\tilde{S}=\frac{1}{2\pi}\int d^2z(k\partial\rho\bar{\partial}\rho+\beta\bar{\partial}\tilde{\gamma}+\bar{\beta}\partial\tilde{\bar{\gamma}}-\frac{\beta\bar{\beta}}{k}e^{-2\rho}).
\end{equation}
The equations of motion of $\beta$ and $\bar{\beta}$ lead to
\begin{equation}
\beta=k\partial\tilde{\bar{\gamma}} e^{2\rho},\quad \bar{\beta}=k\bar{\partial}\tilde{\gamma} e^{2\rho}.
\end{equation}
The Noether currents generating translational symmetry along the lightcone coordinates $\tilde{\gamma}$ and $\tilde{\bar \gamma}$ (\ref{noecurrent})
 in Poincar\'e AdS$_3$ are given by
\begin{equation}\label{tildej}
\tilde{j}^{\bar{z}}_{(\tilde{\gamma})}
=ke^{2\rho}\partial\tilde{\bar{\gamma}}=\beta,\quad \tilde{j}^{z}_{(\tilde{\bar{\gamma}})}=ke^{2\rho}\bar{\partial}{\tilde{\gamma}}=\bar\beta.
\end{equation}

The spectrum of string theory on $AdS_3$ with NS-NS background has been analyzed in \cite{Maldacena:2000hw}, which can be constructed from representations of the $SL(2,\mathbb{R})$ current algebra, 
labelled by $(w; j, \bar j; h,\bar h)$.
The parameter $j$ labels the Casimier of the global $SL(2,\mathbb R)$ algebra.  $h$ and $\bar h$ are the spacetime conformal weights in the dual CFT$_2$, which  are also related to the left and right moving energies in global AdS$_3$. The spectral flow parameter $w$ can be understood as the winding number of a string around the (contractable) spatial circle of global AdS$_3$, and holographically it is related to the twisting number $w$ in the twisted sector. 
The full spectrum of AdS$_3$ consists of the short string sector and long string sector, where the short strings refer to 
the discrete representations $\mathcal D_j$ of $SL(2,\mathbb{R})$ and its spectral flow images $\mathcal D^w_j$, and the long strings refer to the spectral flowed continuous representations  $\mathcal C^w_{j,\alpha}$, where $0\leq\alpha<1$, $j={1\over 2}+is$ with real $s$.
In addition, 
physical states have to satisfy the Virasoro constraint,
\begin{equation}\label{virbeforeflow}
\Delta+\Delta_{rest}=1,\quad  \bar{\Delta}+\bar{\Delta}_{rest}=1  
\end{equation}
where $\Delta$($\bar{\Delta}$) denotes the worldsheet conformal weight of the zero modes of AdS$_3$, and $\Delta_{rest}$($\bar{\Delta}_{rest}$) denotes the worldsheet conformal weight of the oscillating modes as well as the internal spacetime ${\cal N}$. The  worldsheet conformal weight $\Delta$ and $\bar{\Delta}$ can be expressed in terms of the data of the representation as \footnote{Note that sign convention of $w$ here   differs from that of \cite{Apolo:2019zai} by a minus sign.}
\eq{
\Delta=-{j(j-1)\over k-2}-w (h+w {k\over 4}),\quad \bar{\Delta}&=-{j(j-1)\over k-2}-w (\bar{h}+w {k\over 4})\,.\label{spectralflow}} 
Note that the above spectrum was originally derived in global AdS$_3$ by a spectral flow transformation that introduces winding along the spatial circle \cite{Maldacena:2000hw}. Subsequently
 the dependence of worldsheet conformal weight $\Delta $ in \eqref{spectralflow} on the spacetime conformal weight $h$ is introduced because $h$ is related to the momentum of the circle along which the strings winds on.
Nonetheless, the relation \eqref{spectralflow} remains applicable in Poincar\'{e} AdS$_3$, with the understanding that $h$ is still the spacetime conformal weight, but no longer denotes momentum.  From the perspective of the dual CFT, this simply corresponds to the state-operator correspondence. The integer-valued spectral flow parameter $w$ still corresponds to the twisting number in the dual field theory. The fact that \eqref{spectralflow} continues to hold in Poincar\'{e} AdS$_3$ is of great significance for the rest of this paper, where all discussions will be conducted in Poincar\'{e} coordinates.

\subsection{Momentum Dependent Weight After TsT}

In this section, we will derive an equation for the conformal weight after the TsT transformation. The main idea is to relate the TsT transformation to a momentum-dependent spectral flow. In the following we will follow the steps outlined in \cite{Apolo:2019zai} and the conventions therein. 

In order to do so, we introduce three sets of variables, $\tilde{X}$ as the undeformed AdS$_3$ coordinates, $X$ as those after the TsT transformation, and a set of auxilliary coordinates $\hat{X}$ defined by 
\begin{equation}\label{TST}
\partial_a \hat{X}^n=\partial_a X^n-2\tilde{\lambda}\epsilon^{nn^{\prime}}\epsilon_{ab}{j}^b_{(n^{\prime})},
  \end{equation}
where $n$ and $n^{\prime}$ in $\epsilon^{nn^{\prime}}$ represent the two directions along which we perform the TsT transformation. Hatted coordinates in other directions are the same as the unhatted ones, i.e. ${\hat X}^\mu=X^\mu$. 
By using the non-local coordinate transformation \eqref{TST}, the equation of motion and Virasoro constraint satisfied by the coordinates $X^\mu$ after the TsT transformation are equivalent to those before the TsT transformation satisfied by the non-local coordinates $\hat X$ \cite{Frolov:2005dj, Rashkov:2005mi, Alday:2005ww}. 
In particular, the currents are related by \cite{Alday:2005ww},
\begin{equation}
\hat{j}_{m}(\hat{X}^n)=j_{m}(X^n),\quad \hat{j}_{\bar{m}}(\hat{X^n})=j_{\bar{m}}(X^n)
\end{equation}
where $\hat{j}$ and $j$ takes a similar form as (\ref{noecurrent}) but with the background fields and coordinates changed correspondingly. 

In Poincar\'{e} coordinates \eqref{metricandbform}, we consider a TsT transformation along the lightcone coordinates $\tilde{\gamma}$ and $\tilde{\bar{\gamma}}$ with 
$\epsilon^{\tilde{\gamma}\tilde{\bar{\gamma}}}=-1$.
Using the expression (\ref{TST}), we can write down the corresponding relations between hatted variables and those without hats as follows
\eq{\bar{\partial} \hat{\gamma}=\bar{\partial}{\gamma}-2\tilde{\lambda} \bar{\beta},\nonumber\\
\partial\hat{\bar{\gamma}}=\partial\bar{\gamma}-2\tilde{\lambda}\beta. \label{redef}}
 Formulas (\ref{redef}) map solutions of the TsT-transformed background to new solutions with momentum-dependent boundary conditions, 
\bea\label{bdgamma}
\hat{\gamma}(\sigma+2\pi)-\hat{\gamma}(\sigma)&=&\gamma(\sigma+2\pi)-\gamma(\sigma)+{4\pi\tilde{\lambda}}\bar{p}.
\\
\hat{\bar{\gamma}}(\sigma+2\pi)-\hat{\bar{\gamma}}(\sigma)&=&\bar{\gamma}(\sigma+2\pi)-\bar{\gamma}(\sigma)-{4\pi\tilde{\lambda}} p.\nonumber
\eea
where $p$ and $\bar{p}$ represent the momentum along the $\gamma$ and $\bar{\gamma}$ direction after TsT deformation respectively,
\begin{equation}
p=\frac{1}{2\pi}\oint^{2\pi}_0 \beta,\quad \bar{p}=\frac{1}{2\pi}\oint^{2\pi}_0 \bar{\beta}.
\end{equation}
Note that in \eqref{bdgamma} the boundary conditions of $\gamma$ and $\bar \gamma$ are not specified explicitly. We will come to this point later. 

The change in the boundary conditions (\ref{bdgamma}) can be realized by adding a  term linear in the worldsheet coordinates $z$ and $\bar{z}$ to the tilded coordinate,
\begin{equation}\label{rwwp}
\hat{\gamma}=\tilde{\gamma}+2\tilde{\lambda}\bar{p}{z}, \quad \hat{\bar{\gamma}}=\tilde{\bar{\gamma}}+{2\tilde{\lambda}} p\bar{z},
\end{equation}
where $\tilde{\gamma}$ and $\tilde{\bar{\gamma}}$ satisfy the worldsheet equations of motion in Poincar\'{e} AdS$_3$.
 The shift \eqref{rwwp} guarantees that the hatted variables satisfy the same worldsheet equations of motion as the tilded variables do.  
 Comparing \eqref{rwwp} and \eqref{bdgamma}, we learn that $\gamma$ and $\tilde\gamma$ satisfy the same boundary conditions. In this paper, we will choose $\tilde{\gamma}$ in Poincar\'{e} AdS$_3$ to satisfy a twisted boundary condition that corresponds to the representation labelled by integer parameter $w$ as in \eqref{TST}. Note that the $w-$twisting is prior to the TsT transformation, and is not affected by the latter.
The net effect of the TsT transformation is a further twist of the boundary conditions that depends on the TsT parameter $\tilde \lambda$ and the momemtum $p$, 
 which enables us to use a momentum dependent spectral flow to relate the spectrum before and after the TsT transformation. 
 
 In order to derive the spectrum, we need to write down the Virasoro constraints. 
After the TsT transformation, we can write down the stress-energy tensor $
T_{ab}=\frac{2}{\sqrt{-\eta}}\frac{\delta S}{\delta\eta^{ab}}$ in terms of the hatted variables as:
\begin{equation}
T(X)=\hat{T}(\hat X)=-\hat{G}_{\mu\nu}\partial\hat{X}^{\mu}\partial\hat{X}^{\nu}, \quad {\bar T}(X)=\hat{\bar{T}}(\hat{X})=-\hat{G}_{\mu\nu}\bar{\partial}\hat{{X}}^{\mu}\bar{\partial}\hat{{X}}^{\nu},
\end{equation}
where $\hat{G}=\tilde{G}$ is the  metric before the TsT transforamtion. 
The stress-energy tensor before TsT, $\tilde{T}$ and $\tilde{\bar{T}}$, have similar expressions in terms of tilded variables. 
Plugging the metric \eqref{metricandbform} in Poincar\'{e} coordinates into the above expressions, we get the AdS$_3$ part of the  worldsheet stress tensor
\bea
\hat{T}&=&-k(\partial\rho)^2+ke^{2\rho}\partial\hat{\gamma}\partial\hat{\bar{\gamma}}=\tilde{T}-2\tilde{\lambda} \bar{p}\beta\\
\hat{\bar{T}}
&=&-k(\bar\partial\rho)^2+ke^{2\rho}\bar{\partial}\hat{\gamma}\bar{\partial}\hat{\bar{\gamma}}=\tilde{\bar{T}}-2\tilde{\lambda} p\bar{\beta},\nonumber
\eea
where in the second equality we have  used the relation (\ref{rwwp}).
Then from the definition of the Virasoro zero modes $\hat{L}_0=-\frac{1}{2\pi}\oint\hat{\bar T}$ and $\hat{\bar{L}}_0=-\frac{1}{2\pi}\oint\hat{T}$ we get the following relation 
\begin{equation}\label{tstsf}
\hat{L}_0=\tilde{L}_0+{2\tilde{\lambda}} p\bar{p},\quad \hat{\bar{L}}_0=\tilde{\bar{L}}_0+{2\tilde{\lambda}} p\bar{p},
\end{equation}
where $\tilde{L}_0$ is the zero mode of the Virasoro generator before TsT transformation.

Given a representation of $SL(2,\mathbb{R})$ labelled by $(w; j,\bar j; h, \bar h)$,
the expectation value of $\tilde{L}_0$ is given by formula (\ref{spectralflow}).
Such a representation is physical after imposing the Virasoro constraint which now becomes 
\eq{
-{j(j-1)\over k-2}- w (h_{\tilde \lambda }+{k\over 4}w) +{2{\tilde{\lambda} }} p\bar p+ \Delta_{rest} =1,\nonumber\\
 -{\bar j(\bar j-1)\over k-2}-w(\bar h_{\tilde \lambda }+{k\over 4}w) +{2{\tilde{\lambda}}} p\bar p + \bar{\Delta}_{rest}=1.\label{Virafter}
}
where we have fixed the integer-valued winding number $w$, the representation $j$ and also $\Delta_{rest}$, so that the $\tilde \lambda$ dependence is only reflected in $h_{\tilde \lambda }$.
The Virasoro constraints \eqref{virbeforeflow} and \eqref{Virafter} are the on-shell conditions for physical states before and after the TsT transformation, and hence imply the following relation 
\eq{
h_{\tilde \lambda }= h +2{\tilde{\lambda} \over {w} } p \bar p,\quad  \bar h_{\tilde \lambda }= \bar h +2{\tilde{\lambda} \over {w} } p \bar p.
\label{hflow}}
In the special case with $w=1,$ 
we have \eq{
h_{\tilde \lambda }= h +{2{\tilde{\lambda} }} p \bar p,\quad  \bar h_{\tilde \lambda }= \bar h +{2{\tilde{\lambda} }} p \bar p.
\label{hflow2}}
A string state with $w=1$ corresponds to a state in the untwisted sector in the dual field theory ${\rm Sym}^N(\mathcal{M}_\mu)$, the latter of which is also the same as the double-trace $T\bar T$ deformed CFT. 
The shift of the weights (\ref{hflow}) is the key formula of our paper. In the next section, we will  provide a consistency check for this result from OPE between vertex operators, and further apply this result to the calculation of correlation functions. 
 Note that a crucial point in the derivation of \eqref{Virafter} is that we have separated the integer-valued spectral flow parameter $w$ from the momentum-dependent spectral flow \eqref{rwwp} which corresponds to the TsT transformation. 
 The string spectrum on the undeformed background (\ref{spectralflow})  is valid in both global and Poincar\'{e } AdS$_3$, but the relation between the conformal weight $h$ and momentum is only valid in global AdS$_3$. In Poincare AdS$_3$, conformal weight $h$ and momentum $p$ are independent of each other. Similarly, the integer $w$ in Poincar\'{e} corrdiante does not have a simple interpretation as a winding number, but it still corresponds to the twisting number in the dual field theory. A string state satisfying the Virasoro constraint \eqref{Virafter} then corresponds to a state in the $w-$twisted sector with momentum $p,\bar p$, and scaling weight $h_{\tilde\lambda},\,{\bar h}_{\tilde\lambda}$ in the deformed orbifold theory ${\rm Sym}^N(\mathcal{M}_\mu)$.

\section{Correlation Functions From The Worldsheet}
In this section, we will provide a non-perturbative calculation of two-point correlation functions in momentum space after the TsT transformation.  
The basic idea is to construct the vertex operator $\hat{V}$ after the TsT transformation by the momentum-dependent spectral flow, the later of which introduces some dressing factor to the undeformed vertex operators. 
Further using the operator product expansion (OPE), we will re-derive the weight flow formula (\ref{hflow}) and show that the two-point correlation function after TsT transformation can be obtained by replacing the conformal weight $h$ in the undeformed formula by the new weight $h_{\tilde{\lambda}}$.
In the following, we will first derive the non-perturbative correlation functions based on this idea and then discuss its small $\tilde{\lambda}$ expansion, the later of which is found to be consistent with a conformal perturbation calculation on the worldsheet.

\subsection{Non-perturbative Result}

Before the TsT transformation, the long string sector of IIB string theory on AdS$_3\times {\cal N}$ is conjectured to be holographic dual to the symmetric product theory ${\rm Sym}^N(\mathcal{M}_0)$, the latter of which features twisted sectors with twisting number $w$.
According to the holographic dictionary, an operator with spacetime conformal weight $(h,\bar h)$ at location $(x,\bar x)$ in the $w-$twisted sector of ${\rm Sym}^N(\mathcal{M}_\mu)$ corresponds to
a vertex operator on Poincar\'{e} AdS$_3$ \cite{Teschner:1997ft,Zamolodchikov:1986bd} denoted by 
$\tilde V_{j,\bar j,h,\bar h}^{w,\bar w}(z,\bar z;x,\bar x)$ where $(z,\bar z)$ is the coordinate on the worldsheet and $(j,\bar j)$ label the representation of the $SL(2,\mathbb{R})$. As the TsT transformation preserves translational symmetries, it is more convenient to work with vertex operators in the momentum space,  denoted by
$\tilde V_{j,\bar j;h, \bar h}^{w,\bar w}(z,\bar z;p,\bar p)$ where $p$ and $\bar p$ are momenta conjugate to $x$ and $\bar x$. For brevity, henceforth we will often suppress the indices in the right moving sector and only restore them whenever necessary.

The two-point correlation function of the vertex operators have been studied in \cite{Maldacena:2001km,Dei:2021xgh}. Up to a normalization factor, it is given by 
\begin{align} \label{x2pf}
\left\langle 
\tilde{V}^{w}_{j, h}(z_1,x_1)
\tilde{V}^{w}_{j, h}(z_2,x_2)
\right\rangle
\sim 
(x_1-x_2)^{-2h}
(\bar{x}_1-\bar{x}_2)^{-2\bar{h}}
(z_1-z_2)^{-2\Delta}
(\bar{z}_1-\bar{z}_2)^{-2\bar{\Delta}}
\end{align}
Performing Fourier transformation, this two-point correlation function  can be written in the momentum space as 
\begin{equation}\label{undeformed2pf}
\left\langle 
\tilde{V}^{w}_{j, h}(z_1,p)
\tilde{V}^{w}_{j, h}(z_2,-p)
\right\rangle= 
\frac{ \pi 2^{2-4h}\Gamma\left(1-2h\right)}{\Gamma(2h)}
(p\bar{p})^{(2h-1)}\frac{1}{(z_1-z_2)^{2\Delta}}\frac{1}{(\bar{z}_1-\bar{z}_2)^{2\bar{\Delta}}},
\end{equation}
where we have assumed that $h=\bar h$.

As discussed in the previous section, the effect of TsT transformation is encoded in a spectral flow transformation that depends on the momentum. 
Let us denote vertex operators after the TsT transformation by 
$\hat V_{j,\bar j;\hl, \bhl}^{w,\bar w}(z,\bar z;p,\bar p)$, or $\hat V_{j,\hl}^{w}(z;p)$ for brevity.
The boundary conditions (\ref{bdgamma}) imply a branching behavior in the OPE of deformed field $\hat{\gamma}$ and vertex operator $\hat{V}^{w}_{j, h}(z,p)$ 
\begin{equation}\label{twistope}
\hat{\gamma}(z,\bar{z})
\hat{V}^{w}_{j, \hl}(z,p)
\sim {i(2\tilde{\lambda} \bar{p})}\textnormal{ln}(z) \hat{V}^{w}_{j, \hl}(z,p),
\end{equation}
\begin{equation}\label{twistope2}
\hat{\bar{\gamma}}(z,\bar{z})\hat{V}^{w}_{j, \hl}(z,p)\sim {i(2\tilde{\lambda} {p})}\textnormal{ln}(\bar{z}) \hat{V}^{w}_{j, \hl}(z,p).
\end{equation} 
In the following we will construct a vertex operator that satisfies the  OPE \eqref{twistope} and \eqref{twistope2}. 

The field redefinition \eqref{redef} can be integrated by introducing two bosonic fields:
\begin{equation}\label{deform1}
\hat{\gamma}=\gamma-2\tilde{\lambda} \mu_+,\quad \hat{\bar{\gamma}}=\bar{\gamma}-{2\tilde{\lambda}} \mu_-,
\end{equation}
where $\mu_\pm$ are defined by
\begin{equation}\label{mu}
\bar{\beta}=\bar{\partial}\mu_+, \quad \beta=\partial \mu_-.
\end{equation}
As we approach the boundary of AdS$_3$, $\rho$ becomes large and the action \eqref{aaction} becomes that of the $\beta,\gamma$ system together with a free scalar field, so that we can use the following approximation of OPE, 
\bea
\beta(z) \hat{\gamma}(0)&\sim& -\frac{1}{z},
\eea
Then we can get the OPE between $\mu_-$ and $\hat{\gamma}$ as
\begin{equation}\label{eq:OPE-mu-gamma}
\mu_-(z)\hat{\gamma}(0)\sim -\textnormal{ln}(z).
\end{equation}
A similar expression also holds for $\mu_+$ and $\hat{\bar{\gamma}}$. 
By using the OPEs , we can construct the deformed vertex operators satisfying (\ref{twistope}) and (\ref{twistope2}) by applying a dressing factor to the undeformed vertex operator
\begin{equation} \label{eqn:vp_tstq}
\hat{V}^{w}_{j, \hl}(z,p)=e^{{-i2\tilde{\lambda} {p} \mu_+}} e^{{-2i\tilde{\lambda} \bar{p} \mu_-}}\tilde V^{w}_{j, \hl}(z,p)
\end{equation}
 where the explicit form of $h_{\tilde{\lambda}}$ will be derived soon. 
Before the deformation the conformal weights are on-shell in the sense that they are not arbitrary but determined by the Virasoro constraints. To construct physical operators after the TsT transformation, however, we have to start from an off-shell vertex operator with arbitrary conformal weights weights $(\hl, {\bar h}_{\tilde \lambda } )$ and then fix it by physical conditions after the TsT transformation. 
 In the following, we will show that the dressed vertex operators \eqref{eqn:vp_tstq} can be used to reproduce the weight flow equation (\ref{hflow}), and provide a non-perturbative computation of two-point correlation functions in momentum space. 

Using the OPE \eqref{eq:OPE-mu-gamma} and the momentum space correlator  \eqref{undeformed2pf} with weights $(h_{\tilde \lambda }, {\bar h}_{\tilde \lambda })$, we find the two-point function of deformed operators \eqref{eqn:vp_tstq} is very similar to the undeformed one \eqref{undeformed2pf} 
\begin{equation}
\left\langle 
\hat{V}^{w}_{j, h_{\tilde \lambda}}(z_1, p)
\hat{V}^{w}_{j, h_{\tilde \lambda}}(z_2,-p)
\right\rangle=
\frac{ \pi 2^{2-4h_{\tilde \lambda }}\Gamma\left(1-2h_{\tilde \lambda }\right)}{\Gamma(2h_{\tilde \lambda })}
(p\bar{p})^{(2h_{\tilde \lambda }-1)}\frac{1}{(z_1-z_2)^{2\hat{\Delta}}}\frac{1}{(\bar{z}_1-\bar{z}_2)^{2\hat{\bar \Delta}}},
\end{equation}
except that the relation between the weight $h_{\tilde \lambda }$ and the worldsheet conformal weight $\hat \Delta$ is modified to 
\begin{align} \label{eqn:delta_tstq}
    \hat{\Delta} &= -{j(j-1)\over k-2}-w (h_{\tilde \lambda }+{k\over 4}w) +{2{\tilde{\lambda} }} p\bar p, \nonumber \\
    \hat{\bar\Delta}&= -{j(j-1)\over k-2}-w (\bar h_{\tilde \lambda }+{k\over 4}w) +{2{\tilde{\lambda}}} p\bar p .
\end{align}
On the other hand, the dependence in the worldsheet coordinates $(z,\bar z)$ in a physical vertex operator is fixed by conformal invariance. This means that the worldsheet conformal dimenison $\hat\Delta$ should not change under the TsT transformation. In other words, the value of $\hat{\Delta}$ which depends on $ h_{\tilde \lambda }$ equals to the value of $\Delta$ which depends on the undeformed weight $h$, namely 
\be
\hat{\Delta}( h_{\tilde \lambda })=\Delta(h)
\ee
which after plugging in \eqref{eqn:delta_tstq}  and \eqref{spectralflow}, is nothing but the weight flow equation \eqref{hflow} which we reproduce here for convenience 
\eq{
h_{\tilde \lambda }= h +2{\tilde{\lambda} \over {w} } p \bar p,\quad  \bar h_{\tilde \lambda }= \bar h +2{\tilde{\lambda} \over {w} } p \bar p .
\label{hflowsec4}}
We thus conclude that 
the momentum space correlation function after the TsT tansformation will be the same as the one before the TsT transformation with the $z$ dependence factor kept intact, but with the replacement $h\to h_{\tilde \lambda }$ in the momentum-dependent part. More explicitly, 
\begin{align} \label{eqn:deformed_2pt_p}
 \hspace{0.1cm}
\left\langle 
 \hat{V}^{w}_{j_1, \hl}(z_1, p) \,   
 \hat{V}^{w}_{j_2, \hl}(z_2,-p) 
 \right\rangle_{\tilde{\lambda}}  =
 \frac{ \pi 2^{2-4(h+\frac{2\tilde{\lambda} p\bar{p}}{\omega})}\Gamma\left(1-2(h+\frac{2\tilde{\lambda} p\bar{p}}{\omega})\right)}{\Gamma(2(h+\frac{2\tilde{\lambda} p\bar{p}}{\omega}))}
 (p\bar{p})^{2h-1+\frac{4\tilde{\lambda} p\bar{p}}{\omega}} |z_1-z_2|^{-4\Delta}
\end{align}
where we have used the weight flow equation \eqref{hflowsec4}.
{After integrating out the dependence on $z_1$ and $z_2$, up to some extra normalization factors which are not relevant here}, the final form of two-point function in momentum space can be written as
\begin{equation}\label{mastereqn}
G_2^{(\lambda)}(p)= \frac{ \pi 2^{2-4(h+\frac{2\tilde{\lambda} p\bar{p}}{\omega})}\Gamma\left(1-2(h+\frac{2\tilde{\lambda} p\bar{p}}{\omega})\right)}{\Gamma(2(h+\frac{2\tilde{\lambda} p\bar{p}}{\omega}))}
 (p\bar{p})^{2h-1+\frac{4\tilde{\lambda} p\bar{p}}{\omega}}.
\end{equation}

The non-perturbative result of two point correlation function \eqref{mastereqn} is the main result of this paper. Note that our result  holds for the long string sector with non-zero $w$, for which a holographic description in terms of symmetric product theory has been proposed so that a holographic computation in the single-trace $T\bar T$ theory is expected. Correlation functions for the short strings with $w=0$ were discussed in \cite{Chakraborty:2020yka, Giribet:2017imm, Chakraborty:2022dgm}, the holographic interpretation of which is yet to be understood. The fact that the short string sector is holographically more complicated than the long string sector is inherited from the undeformed AdS$_3$/CFT$_2$, where a deformation to the symmetric product theory is necessary \cite{Eberhardt:2021vsx,Giribet:2017imm}.

In the following,
we will test this result by comparing with the first-order perturbative calculation for small $\tilde{\lambda}$.
 Furthermore, we will also see in the next section that this non-perturbative result matches the double-trace CFT non-perturbative result when we take $\omega=1$.
\paragraph{Small $\tilde{\lambda}$ Expansion of the Non-perturbative Result}

We can test (\ref{mastereqn}) by expanding it with respect to small $\tilde{\lambda}$, and then compare it with perturbative result. Expanding the two-point function \eqref{mastereqn} up to the first order of $\tilde{\lambda}$, we get 
\begin{equation}{\label{npexpansion}}
  G_{2}^{(\tilde\lambda)}(p)=G_{2}^{(0)}(p) \left[1-\frac{4\tilde{\lambda}p\bar{p}}{\omega}(\textnormal{log}(4)-\textnormal{log}(p\bar{p})+\psi^{(0)}(1-2h)+\psi^{(0)}(2h))\right] +O(\tilde\lambda^2) 
\end{equation}
where $G_{2}^{(0)}(p)$ is the undeformed two-point function in momentum space
\begin{equation}\label{G0p} 
    G_{2}^{(0)}(p)=\frac{\pi2^{2-4h}\Gamma(1-2h)}{\Gamma(2h)}(p\bar{p})^{2h-1}.
\end{equation}
In the next subsection, we will compare this result with the first-order calculation, which provides a non-trivial test on our non-perturbative result.

\subsection{Perturbative Calculations}

In order to do the perturbative calculation of the correlation function on the AdS$_3$ side, we need to first find the first-order correction of worldsheet action due to the TsT transformation. Following the method in \cite{Apolo:2019zai}, the action after TsT can be written as
\begin{equation}
 S=\frac{1}{2\pi}\int d^2z M_{\mu\nu}\partial X^{\mu}\bar{\partial}X^{\nu}
\end{equation}
where
\begin{equation}
    M=\tilde{M}(I+{2\tilde{\lambda}}\Gamma \tilde{M}),\quad \Gamma_{\mu\nu}=\delta^1_{\mu}\delta_{\nu}^{\bar{2}}-\delta^{\bar{2}}_{\mu}\delta^1_{\nu}.
\end{equation}
For the background \eqref{metricandbform}, the worldsheet action after TsT transformation becomes
\begin{equation}
    S=\frac{k}{2\pi}\int d^2 z(\frac{e^{2\rho}}{1+{2\tilde{\lambda}ke^{2\rho}}}\partial \bar{\gamma}\bar{\partial}\gamma+\partial\rho\bar{\partial}\rho).
\end{equation}
When $\tilde{\lambda}$ is small, we can get
\begin{equation}
S\sim S_0-\frac{\tilde{\lambda}}{\pi }\int d^2z (ke^{2\rho})^2\partial\bar{\gamma}\bar{\partial}\gamma = S_0-\frac{\tilde{\lambda}}{\pi}\int d^2z  j^- \bar j^- ,
\end{equation}
which is evidently a current-current deformation on the worldsheet. 

As will be explained later, the corrections to the correlation function come not only from the change of action, but also from the change of operators. In general, the first-order perturbative expansion of correlation function can be written as \footnote{Here we are using Euclidean signature on the string world sheet.}:
\begin{eqnarray}\label{1storderep}
&&
\left\langle 
 \tilde{V}^{w}_{j_1, h_{\tilde\lambda}}(x_1, z_1) \,   
 \tilde{V}^{w}_{j_2, h_{\tilde\lambda}}(x_2,z_2) 
 \right\rangle_{\tilde\lambda}-\left\langle 
 \tilde{V}^{w}_{j_1, h}(x_1, z_1) \,   
 \tilde{V}^{w}_{j_2, h}(x_2,z_2) 
 \right\rangle_{0}\\
 &=&
 \frac{\tilde\lambda}{\pi}\int d^2z\left\langle 
 j^- \bar j^-
 \tilde{V}^{w}_{j_1, h}(x_1,z_1) \,   
 \tilde{V}^{w}_{j_2, h}(x_2,z_2) 
 \right\rangle_{0}+\left\langle 
  \tilde{V}^{w}_{j_1, h_{\tilde\lambda}}(x_1,z_1) \,   
 \tilde{V}^{w}_{j_2, h_{\tilde \lambda }}(x_2,z_2) 
 \right\rangle_{1}+O\left(\tilde{\lambda}^2\right),
\nonumber\end{eqnarray}
where the fist term comes from the deformed action, and the second term comes from the modified vertex operators. The Ward identity on the worldsheet is \cite{Maldacena:2000hw, Eberhardt:2019ywk}
\begin{equation}\label{jvope}
      j^-(z') \tilde{V}_{j,h}^w(z;x) = \sum_{m=1}^{w} \frac{(j^-_{m} \tilde{V}_{j,h}^w)(z;x)}{(z-z')^{m+1}}+\frac{\partial_x \tilde{V}_{j,h}^w(z;x)}{z-z'}+O(1)\;,
\end{equation}
where $j^-_{m}$ is the $m$-th mode of $j^-$
\footnote{Note that $j^-$ used here is $j^+$ in \cite{Maldacena:2000hw}.}.
The first-order correction to the correlation functions due to the $j^-{\bar j}^-$ deformation is 
\begin{align}\label{jjvv}
&\int dz^2 \left\langle 
 j^-(z) \bar j^-(\bar z)
 \tilde{
 V}^{w}_{j_1, h}(x_1,z_1) \,   
 \tilde{V}^{w}_{j_2, h}(x_2,z_2)
 \right\rangle_{0} \\ 
 &\hspace{1cm} = 
 \frac{4h^2}{|x_{12}|^2}
 \int dz^2 \frac{|z_{12}|^2}{|z-z_1|^2 |z-z_2|^2} \left\langle 
 \tilde{
 V}^{w}_{j_1, h}(x_1,z_1) \,   
\tilde{V}^{w}_{j_2, h}(x_2,z_2)
 \right\rangle_{0}.
 \nonumber 
\end{align}
Note here, even though there are non-trivial terms from descendants in (\ref{jvope}), their contribution to (\ref{jjvv}) is actually zero in our case and we refer the details of this calculation to Appendix \ref{App1}. 
By using dimensional regularization, with $d=2+\tilde{\epsilon}$, the integral of (\ref{jjvv}) can be carried out so that we have  
\begin{equation}
\begin{aligned}\label{jjVV}
&\frac{\tilde{\lambda}}{\pi}\int d^2z\left\langle 
 j^{-}(z) \bar{j}^-(\bar{z})
 \tilde{V}^{w}_{j, h}(x_1, z_1) \,   
 \tilde{V}^{w}_{j, h}(x_2, z_2)
 \right\rangle_{0} \\
 \sim& \ 16\tilde{\lambda} h^2 x_{12}^{-2h-1} \bar{x}_{12}^{-2\bar{h}-1} z_{12}^{-2\Delta} \bar{z}_{12}^{-2\bar{\Delta}} 
\left( \frac{2}{\tilde{\epsilon}}+\log{|z_{12}|^2}+\gamma+\log{\pi}+O(\tilde{\epsilon}) \right),
\end{aligned}
\end{equation}
where we have used the explicit expression of the undeformed two-point function in position space, 
\be
\left\langle 
 \tilde{V}^{w}_{j, h}(x_1, z_1) \,   
 \tilde{V}^{w}_{j, h}(x_2, z_2)
 \right\rangle_{0}=x_{12}^{-2h} \bar{x}_{12}^{-2\bar{h}} z_{12}^{-2\Delta} \bar{z}_{12}^{-2\bar{\Delta}} .\label{G20x}
\ee
 
 Note that the $\log{|z_{12}|^2}$ term in \eqref{jjVV} 
violates conformal symmetry on the worldsheet, and has to be cancelled by a renormalization of the vertex operator, reason why the second term in \eqref{1storderep} is needed.
The appearance of $\log{|z_{12}|^2}$ can also be understood as effectively changing the worldsheet conformal weight $\Delta$ and $\bar{\Delta}$. However, the latter must be fixed by the Virasoro constraints, which necessitates a renormalization of the conformal weight of the vertex operator. Under a shift of the weight $h\to h+\delta h$, the two-point function \eqref{G20x} acquires a correction which, up to the first order of $\delta h$, is given by
\be
  \left\langle 
  \tilde{V}^{w}_{j_1, h_{\tilde \lambda }}(x_1,z_1) \,   
 \tilde{V}^{w}_{j_2, h_{\tilde \lambda }}(x_2,z_2) 
 \right\rangle_{1}\sim -x_{12}^{-2h} \bar{x}_{12}^{-2\bar{h}} z_{12}^{-2\Delta} \bar{z}_{12}^{-2\bar{\Delta}} (2\delta h \log |x_{12}|^2+ 2 \delta \Delta\log |z_{12}|^2 ),
 \ee
where \be \delta \Delta=-w \delta h \label{ddelta}\ee is the change of the worldsheet dimension caused by the shift of $h$, as can be seen from the explicit relation \eqref{spectralflow}. 
To cancel the $\log |z_{12}|^2$ term in \eqref{jjVV}, we have to choose 
\be \delta h=-{8\tilde{\lambda} h^2\over w |x_{12}|^2},\label{hflowx}
\ee
which can be interpreted as the infinitesimal version of the weight flow \eqref{hflow} in position space. 
The weight flow \eqref{hflowx} guarantees that the $z$ dependence in the correction \eqref{1storderep} remains the same as in the undeformed correlator \eqref{eqn:deformed_2pt_p}. The first-order correction \eqref{1storderep} now becomes 
\be\label{1storder}
\begin{aligned}
16 \tilde{\lambda} h^2 x_{12}^{-2h-1} \bar{x}_{12}^{-2\bar{h}-1} z_{12}^{-2\Delta} \bar{z}_{12}^{-2\bar{\Delta}}\left({1\over\omega}\textnormal{log}|x_{12}|^2 + \frac{2}{\tilde{\epsilon}}+\gamma+\log{\pi}+O(\tilde{\epsilon})\right).
\end{aligned}
\ee
Note that the physical result should be independent of regularization parameter $\tilde{\epsilon}$, which can be removed by further renormalizing the vertex operator,
\begin{equation}
\tilde{{V}}^w_{j,h}\to\tilde{V}^w_{j,h}-2\tilde{\lambda}\left( \frac{2}{\tilde{\epsilon}}+\gamma+\log{\pi}\right)\nabla_x\nabla_{\bar{x}}\tilde{V}^w_{j,h}  
\end{equation}
Then by combining (\ref{1storderep}) and (\ref{1storder}), we obtain the final result
\begin{equation}
\begin{aligned}\label{pertxTsT}
{\left\langle 
\tilde{{V}}^{w}_{j_1, h_{\tilde \lambda }}(x_1,z_1) \,   
 \tilde{{V}}^{w}_{j_2, h_{\tilde \lambda }}(x_2,z_2) 
 \right\rangle_{\tilde \lambda}\over \langle\tilde{{V}}^{w}_{j_1, h_{\tilde \lambda }}(x_1,z_1) \,   
 \tilde{{V}}^{w}_{j_2, h_{\tilde \lambda }}(x_2,z_2)
 \rangle_0 } \; = \;
 \Big(1+ \frac{16\tilde{\lambda} h^2}{\omega|x_{12}|^2}\log|x_{12}|^2+{\cal O}(\tilde{\lambda}^2)\Big).
 \end{aligned}
\end{equation}

\paragraph{Momentum space}
To check the consistency of our result, it is necessary to compare the first-order perturbative result with the small $\tilde\lambda$ expansion of the non-perturbative result. Since the non-perturbative result is in momentum space, let us calculate our first-order correction in momentum space directly. Using the coordinate space OPE (\ref{jvope}) and the fact that contributions from descendant fields vanish, we obtain the first-order correction from the deformed action to two-point functions as
\begin{equation}
    \begin{aligned}\label{p1loop}
    &\frac{\tilde{\lambda}}{\pi}\int dz^{2}\left\langle j^{-}(p,z)\bar{j}^{-}(\bar{p},\bar{z})\tilde{V}_{j_{1},h}^{w}(p,z_{1})\,\tilde{V}_{j_{2},h}^{w}(-p,z_{2})\right\rangle _{0}\\
    =&-4\tilde{\lambda}G_{2}^{(0)}(p)p\bar{p}z_{12}^{-2\Delta}\bar{z}_{12}^{-2\bar{\Delta}}\left(\frac{2}{\tilde{\epsilon}}+\log|z_{12}|^{2}+\gamma+\log\pi+O(\tilde{\epsilon})\right),
    \end{aligned}
\end{equation}
where $G_{2}^{(0)}(p)$ is the undeformed two-point function in momentum space \eqref{G0p}.
Just like the  calculation in coordinate space, to ensure conformal symmetry on the worldsheet, we have to renormalize the conformal weight, which leads to an additional term 
\begin{equation}\label{p1ren}
    \begin{aligned}
  &\left\langle \tilde{V}_{j_{1},h_{\tilde \lambda }}^{w}(p,z_{1})\,\tilde{V}_{j_{2},h_{\tilde \lambda }}^{w}(-p,z_{2})\right\rangle _{1}\\
  =&-G_{2}^{(0)}(p)z_{12}^{-2\Delta}\bar{z}_{12}^{-2\bar{\Delta}}\Big(2\delta \Delta\log |z_{12}|^{2}+2\delta h\big(-\log p\bar{p}+\log4+\psi^{(0)}(1-2h)+\psi^{(0)}(2h)\big)\Big),
    \end{aligned}
\end{equation}
where $2\delta \Delta=-4\tilde{\lambda}p\bar{p}$ is required to cancel the $\log |z_{12}|^{2}$ in \eqref{p1loop}. Similar to the discussion in postion space, after using the relation $\delta \Delta=-w\delta h$ as determined by \eqref{spectralflow}, we get an infinitesimal flow of the weight \be \delta h=\frac{2\tilde{\lambda}}{\omega}p\bar{p}.\ee 
This is nothing but the non-perturbative weight flow equation \eqref{hflow}. 

As what we did in position space, here we can apply a renormalization process to get rid of the part involving $\left(\frac{2}{\tilde{\epsilon}}+\gamma+\log\pi+O(\tilde{\epsilon})\right)$. 
After doing this, the sum of \eqref{p1loop} and \eqref{p1ren} gives rise to the final first-order correction to the momentum space two-point function, and the result matches the small $\tilde{\lambda}$ expansion of the non-perturbative result (\ref{npexpansion}).

\section{Correlation Functions From $T\bar T$ }
In this section, we first briefly review the symmetric orbifold CFT, and its single-trace $T\bar{T}$ deformation. We will compute the correction to two-point functions up  to the linearized order in the deformation parameter using Ward identity, and study the non-perturbative Callan-Symmanzik equation in the double-trace $T\bar{T}$  deformation. We will compare both the perturbative and non-perturbative results to those derived from the string worldsheet.

\subsection{Conformal Perturbation}
In this subsection, we briefly review some basic facts of symmetric orbifold CFT ${\rm Sym}^N(\mathcal{M}_0)$, and then calculate the single-trace $T\bar{T}$ deformed two-point function of the twist operator using conformal perturbation.

The orbifold CFT ${\rm Sym}^N(\mathcal{M}_0)$ is defined by considering $N$ copies of CFT with a quotient action.
The field $\Phi_I$, $I=1,\cdots,N$, of each copy satisfy the equivalence relation
$\Phi_I\sim \Phi_{g(I)}$ for any $g\in S_N$. 
The twist field $\sigma_g(x)$ is introduced to create the twist vacua, such that the filed $\Phi_I$ obeys 
\be
\Phi_I(e^{2\pi i}(x-x_1))\sigma_g (x_1)=\Phi_{g(I)}(x-x_1)\sigma_g(x_1),
\ee
where $x$ represents coordinates of the two-dimensional complex plane $ds^2=dxd\bar{x}$. See \cite{Lunin:2001ew,Roumpedakis:2018tdb} for more details of the orbifold CFT.

The field $\Phi_I(x)$ is not single-valued in the $x$-plane. We can replace the $n$ fields by a single field $\Phi(z)$ living on $x$-plane's covering space with coordinate $t$\cite{Lunin:2001ew}. Passing to the covering space, it is easy to find the conformal dimension of the length $n$ twisted operator $\sigma_n$ is $ h_n=\frac{c}{24}(n-\frac{1}{n})$. 
The gauge invariant correlator of twist operators should be the following one
\be
\langle \sigma_{g_1} \sigma_{g_2}\dots \sigma_{g_n}\rangle,
\ee
with $g_1g_2\dots g_n=1$. For bosonic string with $k=3$, it is conjectured 
that the twist operator corresponds to the vertex operator $\tilde{V}^{n}_{j, h}$ with $j={1\over2}, h=h_n$, which represents the ground state of a single string with winding number $n$ \cite{Eberhardt:2021vsx}. For general value of $k$, a general vertex operator $\tilde{V}^{n}_{j, h}$ corresponds to $\sigma_{n,\alpha}$, which is the twisted field dressed by a local operator $O_{\alpha}$ of the seed theory.

 To derive the Ward identity of the stress tensor $\langle T_I(x)\sigma_{g_1}\sigma_{g_2}\dots\sigma_{g_n}\rangle$, it is convenient to move to the covering space. Since the one point function of the stress tensor in the covering space is always zero, one finds the only non-trivial contribution is due to the Schwarzian of the coordinate transformation
\be
\frac{\langle T_I(x)\sigma_{g_1}\sigma_{g_2}\dots\sigma_{g_n}\rangle}{\langle\sigma_{g_1}\sigma_{g_2}\dots\sigma_{g_n}\rangle}=\frac{c}{12}\{t_I,x\},
\ee
where $t_I$ is the inverse map of $x(t)$ which corresponds to the coordinates where $\Phi_I$ lives. See 
\cite{Roumpedakis:2018tdb} for the case of the free field symmetric orbifold CFT. For the part relevant to our calculation, the Ward identity of $T\bar{T}$ operator is 
\be\label{eq:Ward-CFT}
\frac{\langle T_I\bar{T}_I(x)\sigma_{g_1}\sigma_{g_2}\dots\sigma_{g_n}\rangle}{\langle\sigma_{g_1}\sigma_{g_2}\dots\sigma_{g_n}\rangle}=(\frac{c}{12})^2\{t_I,x\}\{\bar{t}_I,\bar{x}\}.
\ee
Let us consider the single-trace $T\bar{T}$ deformation of the symmetric orbifold CFT defined by
\be
\frac{\partial S_\mu}{\delta \mu}={-\frac{1}{\pi}\int d^2x}\sum_{I=1}^N (T_I\bar{T}_I)_\mu,
\ee
where $\mu$ is the deformation parameter. Let us consider the single-trace $T\bar{T}$ deformation of the two-point function 
\be
\langle \sigma_{(12\cdots n)}(0)\sigma_{(n\cdots 21)}(a)\rangle,
\ee
where the two twisted operators are located at $x=0,a$ respectively. The first-order correction due to the deformation is
\be\label{std1}
{\frac{1}{\pi}}\mu\sum_{I=1}^{n}\int d^{2}x\langle T_{I}\bar{T}_{I}(x)\sigma_{(12\cdots n)}(0)\sigma_{(n\cdots21)}(a)\rangle.
\ee
The covering map in this case is given by \cite{Lunin:2001ew,Lunin:2000yv,Pakman:2009zz}
\be
x=a\frac{t^{n}}{t^{n}-(t-1)^{n}}.
\ee
The inverse functions in this case are given by $t_I$, with $I$ from $1$ to $n$, each representing one sheet of the covering space. We find all the Schwarzians have the same value
\be\label{sch}
\{t_I(x),x\}=\frac{a^2 \left(n^2-1\right)}{2 n^2 x^2 (a-x)^2},
\ee
for each $t_I$. Therefore, we can calculate the first-order correction for each copy and then sum up. The result is
\be{\label{cftperturbative}}
\baa
&{\frac{1}{\pi}}\mu\sum_{I=1}^{n}\int d^{2}x\langle T_{I}\bar{T}_{I}(x)\sigma_{(12\cdots n)}(0)\sigma_{(n\cdots21)}(a)\rangle
\\
=&{\frac{1}{\pi}}\mu(\frac{c}{24})^{2}\frac{1}{n}\left(n-\frac{1}{n}\right)^{2}|a|^{4}\int d^{2}x\frac{1}{|x|^{4}|a-x|^{4}}\langle\sigma_{(12\cdots n)}(0)\sigma_{(n\cdots21)}(a)\rangle\\=&\frac{16\mu h_n^2}{n|a|^{2}}\big(\frac{2}{\tilde{\epsilon}}+\log|a|^{2}+\gamma+\log\pi-5/2+{\cal O}(\epsilon)\big)\langle\sigma_{(12\cdots n)}(0)\sigma_{(n\cdots21)}(a)\rangle
\eaa
\ee
with $h_n=\frac{c}{24}(n-\frac{1}{n})$. 
Here $\tilde{\epsilon}$ is the parameter of the dimensional regularization by replacing 2 to $(2+\tilde{\epsilon})$. The divergence $1/\tilde{\epsilon}$ can be renormalized by using the same procedure in the previous section, which leads to the final result 
\begin{equation}\label{eq:first order-renom}
\begin{aligned}
{\langle\sigma_{(12\cdots n)}(0)\sigma_{(n\cdots21)}(a)\rangle_{\mu} \over\langle\sigma_{(12\cdots n)}(0)\sigma_{(n\cdots21)}(a)\rangle}= 
 \left(1+\frac{16\mu h_{n}^{2}}{n|a|^{2}}\log(|a/\epsilon|^{2})+O(\mu^2)\right),
\end{aligned}
\end{equation}
where $\epsilon$ is the  UV cutoff of the theory and $\langle\sigma_{(12\cdots n)}(0)\sigma_{(n\cdots21)}(a)\rangle$ is the two-point function of the original CFT which has the standard form,
\begin{equation}
\langle\sigma_{(12\cdots n)}(0)\sigma_{(n\cdots21)}(a)\rangle= \frac{C_{CFT}}{|a|^{4h_n}} , 
\end{equation}
where $C_{CFT}$ is a normalization factor.
Note that to obtain the first-order result \eqref{eq:first order-renom}, we have used dimensional regularization followed with a renormalization which eliminates the regulator $\tilde \epsilon$, but reintroduced a cutoff parameter $\epsilon $ by dimensional analysis. Alternatively, the same result (\ref{eq:first order-renom}) can be obtained directly by the point splitting method \cite{Cardy:2019qao,Hirano:2020ppu}, where the parameter $\epsilon$ plays the role of the cutoff of point-splitting. See appendix \ref{sec:2pt-ttbar} for the details of this method. Although the above result of two-point function \eqref{eq:first order-renom} is for the twist operators only, it is plausible to expect the general structure holds for more general primary operators. 

Correlators in the untwisted sector of sing-trace $T\bar T$ deformation with $n=1$ should be the same as
that of the double-trace $T\bar T$ deformation, where the Ward identity for a primary operator can be written as \be
T (x){ O} (y)=\big({h\over (x-y)^2}  +{\p_y\over x-y}\big){ O} (y).\label{warddt}
\ee 
Using \eqref{warddt}, we can calculate the first-order correction to the two-point function, which after renormalization can be written as  \cite{Kraus:2018xrn,Cardy:2019qao}
\begin{equation}\label{eq:first order-gen}
\begin{aligned}
{\langle{O} (0){ O} (a)\rangle_{\mu}\over \langle{O} (0){O} (a)\rangle } = 
 \left(1+\frac{16\mu h^{2}}{|a|^{2}}\log(|a/\epsilon|^{2})+O(\mu^2)\right)
.
\end{aligned}
\end{equation}
Comparing with the double-trace $T\bar{T}$ deformed two-point function, we find the single-trace result \eqref{eq:first order-renom} is the same except for a factor of $1/n$, a structure that typically appears in the twisted sector. 
This provides an additional reason why we expect that the result of \eqref{eq:first order-renom} also holds in the case of generic operators $O$. 

\subsubsection*{Comparison to the TsT calculation}
In order to make comparisons to the results obtained from the worldsheet calculation, let us recall the holographic dictionary \be\label{dictionary}
\ell^2\tilde{\lambda}=\mu,\quad w=n
\ee
which states that the parameter of TsT transformation $\tilde \lambda$ and  the winding number $w$ of long strings on the string theory side 
are identified with the deformation parameter $\mu$ and the length of a cycle in the twisted sector of the symmetric product theory ${\rm Sym}^N(\mathcal{M}_0)$.

Using conformal perturbation theory on the string worldsheet, the two-point correlation function is given by \eqref{pertxTsT}, which we reproduce here for convenience, 
\begin{equation}
\begin{aligned}\label{pertxTsT2}  
{\left\langle 
\tilde{{V}}^{w}_{j_1, h_{\tilde \lambda }}(x_1,z_1) \,   
 \tilde{{V}}^{w}_{j_2, h_{\tilde \lambda }}(x_2,z_2) 
 \right\rangle_{\tilde\lambda }\over \langle\tilde{{V}}^{w}_{j_1, h_{\tilde \lambda }}(x_1,z_1) \,   
 \tilde{{V}}^{w}_{j_2, h_{\tilde \lambda }}(x_2,z_2) 
 \rangle_0}
 = \Big(1+ \frac{16\tilde{\lambda} h^2}{\omega|x_{12}|^2}\log|x_{12}|^2+{\cal O}(\tilde{\lambda}^2)\Big).
 \end{aligned}
\end{equation}
Using the holographic dictionary \eqref{dictionary} and taking $x_{12}=-a$ and $h=h_n$,
we learn that up to this order, the twist operator two-point function \eqref{eq:first order-renom} has the same structure as that of \eqref{pertxTsT2}, for arbitrary value of $w=n$. 
In addition, in the untwisted sector with $n=1$, the perturbative results from both sides, namely \eqref{pertxTsT2} and \eqref{eq:first order-gen}, agree for arbitrary primary operators.  

As the string theory result \eqref{pertxTsT2} is valid for general operators 
with arbitrary $w=n$, to complete the comparison we still need to provide a derivation of the two-point correlators for general primary operators in the twisted sector, which is expected to take a similar form as \eqref{eq:first order-renom}.  We leave this for future study.

\subsection{Non-perturbative Results}
It would be interesting to calculate the non-perturbative two-point function of the single-trace $T\bar{T}$-deformed theory and compare it with the string theory result (\ref{eqn:deformed_2pt_p}). In this paper we focus only on the untwisted sector with $n=1$, where the results of the single-trace deformation align with those of the double-trace $T\bar{T}$-deformed conformal field theory derived in \cite{Cardy:2019qao}. It has been observed that the logarithmic divergence arises not only in the first-order perturbation \cite{Kraus:2018xrn}, but also at all orders. Therefore, it is necessary to introduce a renormalized operator to eliminate the divergence.
In our convention, we need to cancel the logarithmic divergence of the two-point correlation function \eqref{eq:first order-gen} in momentum space $\langle O(q)O(-q)\rangle_\mu\sim  \langle O(q)O(-q)\rangle_0\Big(1+4\mu q\bar{q}\log(q\bar{q}\epsilon^2)+\cdots\Big)$,
and the renormalized operator should be
\begin{equation}
   {\cal O}(q)=e^{-4\mu\log(\Lambda\epsilon)q\bar{q}}{ {O}}(q) 
\end{equation}
with $\mu$ the deformation parameter, $\epsilon$ the point-splitting cut-off, $\Lambda$ renormalization scale and $q$ the momentum. From the condition that the bare correlator does not depend on renormalization scale $\Lambda$, we can get the Callan-Symanzik equation for a renormalized two-point function $\mathcal{C}_2(q_{n},\mu,\Lambda)$ \cite{Cardy:2019qao}
\be
\baa
0&=\Lambda\partial_{\Lambda}\Big[e^{8\mu\log(\Lambda\epsilon)q\bar{q}}\mathcal{C}_2(q,\mu,\Lambda)\Big],
\eaa
\ee
from which we can get,
\begin{equation}
0=(\Lambda\partial_{\Lambda}+8\mu q\bar{q})\mathcal{C}_2(q,\mu,\Lambda).  \label{CSeq}
\end{equation}
On the other hand, from dimensional analysis and the boundary condition $\mathcal{C}_2(q,\mu=0)\sim (q\bar{q})^{2h-1}$, we can also get \footnote{The position version of the Callen-Symanzik equation \eqref{CSeq} is given by eq.(5.11) in \cite{Hirano:2020nwq}, wherein a detailed derivation is also provided.}
\be\label{csequation}
0=\big(\Lambda\partial_{\Lambda}+q\partial_{q}+\bar{q}\partial_{\bar{q}}-2\mu\partial_{\mu}-(4h-2)\big)\mathcal{C}_{2}(q,\mu,\Lambda).
\ee
At fixed $\Lambda$, the above equation becomes 
\begin{equation}
 0=\big(q\partial_{q}+\bar{q}\partial_{\bar{q}}-2\mu\partial_{\mu}-(4h-2)-8\mu q\bar{q}\big)\mathcal{C}_{2}(q,\mu,\Lambda).\label{CSeqf}   
\end{equation}
It is interesting to note that 
\begin{equation}
    \big(q\partial_{q}+\bar{q}\partial_{\bar{q}}-2\mu\partial_{\mu}\big)F(\mu q\bar{q})=0
\end{equation}
ufor any function $F(x).$ Thus a general solution for the Callan-Symanzik equations \eqref{CSeq} and (\ref{CSeqf}) can be written as:
\begin{equation}\label{generalsol}
 F(\mu q\bar{q})(q\bar{q})^{2h-1+4\mu q\bar{q}}\Lambda^{-{8}\mu q\bar{q}},\quad F(0)=1.  
\end{equation}
The coefficient $F(\mu q\bar {q})$ depends on the renormalization scheme and will not be fully determined here. In \cite{Cardy:2019qao}, the function is chosen as $F(\mu q\bar q)=1$. As can be seen momentarily, in order to match the nonperturbative result derived from string theory, however, we need to make a different choice for $F(\mu q\bar q)$ .

\subsubsection*{Comparison to the TsT calculation}
Using the TsT transformation and spectral flow transformation,  we have obtained a non-perturbative result for two-point correlation functions (\ref{mastereqn}). The result with $w=1$ is given by,  
\begin{equation}\label{non-per2}
 \frac{ \pi 2^{2-4h-{8\tilde{\lambda} p\bar{p}}}{\Gamma\left(1-2h-{4\tilde{\lambda} p\bar{p}}\right)}}{\Gamma(2h+{4\tilde{\lambda} p\bar{p}})} (p\bar{p})^{2h-1+{4\tilde{\lambda} p\bar{p}}}.  \end{equation}
The result with $w=n=1$ is to be compared to the untwisted sector of single-trace $T\bar T$ deformed CFT, or equivalently the double-trace $T\bar T$ deformed CFT. 
Using the holographic dictionary \eqref{dictionary}, and identifying the momentum $p=-q$, it is evident that \eqref{non-per2} takes the general form of (\ref{generalsol}), and thus satisfies the Callan-Symanzik equation (\ref{csequation}). 
It will be interesting to further understand the specific choice of the function $F(\mu q\bar q)$ from the $T\bar T$-deformed CFTs.

\section{Conclusion and discussion}
In this paper, we investigate two-point correlation functions in the conjectured correspondence between IIB string theory, which is obtained by applying TsT transformation to the AdS$_3\times \mathcal N$ background with NS-NS background, and the single-trace $T\bar T$ deformed symmetric product theory ${\rm Sym}^N(\mathcal{M}_
\mu)$. Using a momentum-dependent spectral flow, we derive a formula for the flow of the conformal weight (\ref{hflow}) and a non-perturbative result for two-point correlation functions (\ref{mastereqn}). The non-perturbative result is consistent with conformal perturbation on the worldsheet in both position space and momentum space. We demonstrate that, up to the linearized order, the two-point correlation functions of twisted operators with arbitrary twist number $w = n$, and general primary operators in the untwisted sector with $w = n = 1$ of ${\rm Sym}^N(\mathcal{M}_
\mu)$ are both consistent with the string theory result. Finally, the non-perturbative result in the untwisted sector satisfies the Callen-Symanzik equation.

Several interesting questions remain to be explored in future work. To complete the comparison of two-point functions, more work needs to be done from the $T\bar T$ side, including i) a derivation of the arbitrary function $F(\mu q\bar q)$ that appears in the general solution \eqref{generalsol} of the Callen-Symanzik equation, ii) a derivation of the Callen-Symanzik equation in the twisted sector of the deformed orbifold theory ${\rm Sym}^N(\mathcal{M}_\mu)$, and iii) a perturbative calculation of general operators in the twisted sector. On the string theory side, it will be interesting to use our techniques to further compute three and four-point functions. Finally, while this paper focuses on TsT transformations that are dual to single-trace $T\bar T$ deformations, we expect that the worldsheet calculation can be generalized to single-trace $J\bar T$ and $J \bar T + T \bar T + T \bar J$ deformations \cite{Guica:2017lia,Chakraborty:2018vja,Apolo:2018qpq,LeFloch:2019rut,Chakraborty:2019mdf,Frolov:2019xzi,Apolo:2021wcn}.

\section*{Acknowledgments}
We would like to thank Luis Apolo, Pengxiang Hao, Song He, Yunfeng Jiang, Wenxin Lai, Wen-jie Ma, Hao Ouyang and Fengjun Xu for useful discussions. The work of WC is supported by the fellowship of China Postdoctoral Science Foundation NO.2022M720507. The work of HS is supported in part by the Beijing Postdoctoral Research Foundation.
The work of WS is supported by the national key research and development program of China No. 2020YFA0713000. The work of JTW is supported by the fellowship of China Postdoctoral Science Foundation NO.2022M720508. 
WC also would like to thank the Tsinghua Sanya International Mathematics Forum (TSIMF) for the hospitality where part of this work was completed.

\appendix

\section{$j^-\bar{j}^-$  deformation of two-point correlation function} \label{App1}

In this appendix, we will discuss how to derive equation (\ref{jjvv}), i.e. the first-order $j^-\bar{j}^-$ deformation of two-point correlation function from (\ref{jvope}). We will show explicitly that the descendant fields in (\ref{jvope}) will make no contribution in (\ref{jjvv}). 
The OPE between the $SL(2,\mathbb{R})$ currents and vertex operators inserted at $x=\bar x=0$ is given by  
\bea \label{eqn:wardJA}
    j^-(z) V_{j,h}^w(0;0)&\sim& \sum_{m=1}^{w} \frac{(j^-_{m} V_{j,h}^w)(0;0)}{z^{m+1}}+\frac{\partial_x V_{j,h}^w(0;0)}{z}\nn,\\
    j^3(z) V_{j,h}^{w}(0;0)&\sim& \frac{hV_{j,h}^w(0,0)}{z}, \\
   j^+(z) V_{j,h}^w(0;0)&\sim& \mathcal{O}(z^{w-1}) \nonumber.
\eea
vertex operators elsewhere are obtained from those at the origin by evolving it along $x$ and $\bar x$,
\be
V_{j,h}^{w}(x;z_i)=e^{x j^-_0+\bar x {\bar j}^-_0}\Bigl( V_{j,h}^{w}(0;z_i)\Bigr)e^{-x j^-_0-\bar x {\bar j}^-_0}\,.
\ee
Then the analog of \eqref{eqn:wardJA} at arbitrary location is
\be
j^a(\xi)V^w_{j,h}(x,z)=e^{x j^-_0+\bar x {\bar j}^-_0}\Bigl(j^{a(x)}(\xi) V_{j,h}^{w}(0;z)\Bigr)e^{-x j^-_0-\bar x {\bar j}^-_0}\label{WIx}
\ee
with 
\bea\label{jxV}
j^{-(x)}(z)&=&j^{-}(z)\nn,\\
j^{3(x)}(z)&=&j^{3}(z)+xj^{-}(z),\\
j^{+(x)}(z)&=&j^+(z)+2xj^3(z)+j^-(z).\nn
\eea
The first-order correction to the two-point function in the $j\bar{j}$ deformed theory is 
%
\begin{align}\label{jjvvf}
   &\left\langle \bar{j}^-(\bar z) j^-(z) V^{w}_{j,h}(x_1; z_1) \,  V^{w}_{j,h}(x_2; z_2) \right\rangle \\\nonumber
   =& \left \langle  \bar{j}^-(\bar z) \cdot  j^-(z) \cdot V^{w}_{j,h}(x_1; z_1) \,  V^{w}_{j,h}(x_2; z_2) \right \rangle + 
   \left \langle   V^{w}_{j,h}(x_1; z_1) \,  \bar{j}^-(\bar z) \cdot  j^-(z) \cdot V^{w}_{j,h}(x_2; z_2) \right \rangle \\
   & + \left \langle  \bar{j}^-(\bar z) \cdot  V^{w}_{j,h}(x_1; z_1) \,  j^-(z) \cdot V^{w}_{j,h}(x_2; z_2) \right \rangle  + \left \langle  
j^-(z) \cdot V^{w}_{j,h}(x_1; z_1) 
\bar{j}^-(\bar z) \cdot  V^{w}_{j,h}(x_2; z_2)
 \right \rangle \nonumber
\end{align}
where $\cdot$ represents the contraction of $j^{-}$($\bar{j}^{-}$) with the vertex operators.
Using the Ward identity (\ref{WIx}), which in the case of $j^-$ takes a similar form as \eqref{eqn:wardJA},  
the correlation function \eqref{jjvvf} can be written as a sum of (\ref{jjvv}) and additional terms in the following form:
\begin{equation}
 \sum_{\bar m=1}^{w}\sum_{m=1}^{w} \frac{G^1_{\bar{m},m} }{(\bar z- \bar z_1)^{\bar m+1}(z-z_1)^{m+1}};\quad  \sum_{\bar m=1}^{w}\sum_{m=1}^{w} \frac{G^2_{\bar{m},m} }{(\bar z- \bar z_2)^{\bar m+1}(z-z_2)^{m+1}},
\end{equation}
 \begin{equation}
 \sum_{\bar m=1}^{w}\sum_{m=1}^{w} \frac{M_{m, \bar{m}}}{(\bar z- \bar z_2)^{\bar m+1}(z-z_1)^{m+1}};\quad \sum_{\bar m=1}^{w}\sum_{m=1}^{w} \frac{M_{\bar{m}, m}}{(\bar z- \bar z_1)^{\bar m+1}(z-z_2)^{m+1}},
\end{equation}
\begin{equation}
\sum_{\ell=1}^{w} \frac{F^2_{\ell} }{(z-z_2)^{\ell+1}};\quad \sum_{\ell=1}^{w} \frac{F^1_{\ell} }{(z-z_1)^{\ell+1}};\quad\sum_{\ell=1}^{w} \frac{\bar{F^1}_{\ell} }{(\bar z-\bar z_1)^{\ell+1}};\quad\sum_{\ell=1}^{w} \frac{\bar{F^2}_{\ell} }{(\bar z-\bar z_2)^{\ell+1}}.
\end{equation}
 where the $z$/$\bar{z}$-independent coefficients are given by 
 \begin{equation}
     \begin{aligned}
&G^1_{\bar{m},m}=
\left \langle 
\left(\bar{j}^-_{\bar m}j^{-}_{m}V^{w}_{j,h}(x_1; z_1)\right)V^{w}_{j,h}(x_2; z_2)
\right \rangle , \quad 
G^2_{\bar{m},m}=
\left \langle 
V^{w}_{j,h}(x_1; z_1)
\left(\bar{j}^-_{\bar m}j^{-}_{m}V^{w}_{j,h}(x_2; z_2)\right)
\right \rangle , \\
&M_{\bar m,m}=
\left \langle 
\bar{j}^-_{\bar m} V^{w}_{j,h}(x_1; z_1) j^{-}_{m} V^{w}_{j,h}(x_2; z_2)
\right \rangle ,  \quad \quad 
M_{m,\bar m}=
\left \langle 
j^{-}_{m}
V^{w}_{j,h}(x_1; z_1) 
\bar{j}^-_{\bar m}
V^{w}_{j,h}(x_2; z_2)
\right \rangle , 
\\
&F^1_\ell = \langle (j^-_\ell V^{w}_{j,h}(x_1; z_1) ) V^{w}_{j,h}(x_2; z_2) \rangle ,
\qquad \qquad 
F^2_\ell = \langle  V^{w}_{j,h}(x_1; z_1) (j^-_\ell V^{w}_{j,h}(x_2; z_2))   \rangle ,
\\
&\bar{F}^1_\ell = \langle (\bar{j}^-_\ell V^{w}_{j,h}(x_1; z_1) ) V^{w}_{j,h}(x_2; z_2)
\rangle ,
\qquad \qquad 
\bar{F}^2_\ell = \langle  V^{w}_{j,h}(x_1; z_1)  (\bar{j}^-_\ell V^{w}_{j,h}(x_2; z_2) )  \rangle.
     \end{aligned}
 \end{equation}
All the above terms come from the action of non-zero modes of the currents on the vertex operators, i.e. the first term in the right hand side of \eqref{eqn:wardJA}.
In particular, the $G$ terms come from the action of the non-zero modes of ${j}^-$ and $\bar{j}^-$ on the same vertex operators, the $M$ terms from the non-zero modes of ${j}^-$ and $\bar{j}^-$ on different vertex operators. The $F$ terms represent the mixing of non-zero modes with the operation of differential operator. 
Now our task is to show that all the aforementioned  terms are actually zero. 

The G terms always vanish after $z$ integration, as
\begin{align}
&\int d^2z\frac{1}{(z-z_1)^{l_1}(\bar{z}-\bar{z}_1)^{m_1}},\quad l_1\geq 2,\quad m_1\geq 2\\ \nonumber
&=(l_1-1)!(m_1-1)!\partial^{l_1-1}_{z_1}\partial^{m_1-1}_{\bar{z}_1}\int d^2z\frac{1}{|z-z_1|^2}\\\nonumber
&=(l_1-1)!(m_1-1)!\partial^{l_1-1}_{z_1}\partial^{m_1-1}_{\bar{z}_1}\int d^2z\frac{1}{|z|^2}=0
\end{align}
Similarly, all the M terms also vanish. 

For F terms, let's review the trick in \cite{Eberhardt:2019ywk}. One can compute the special combination of the currents in two different ways.  
First, as shown in equation (4.7) in \cite{Eberhardt:2019ywk}, using \eqref{eqn:wardJA} and \eqref{jxV} 
we have 
\begin{align}
\left\langle\Bigl(j^+(z)-2x_j j^3(z)+x_j^2 j^-(z)\Bigr)
V^{w}_{j,h}(x_1; z_1) V^{w}_{j,h}(x_2; z_2) 
\right \rangle=\mathcal{O}((z-z_i)^{w-1})\label{jjvvm1}
\end{align}
On the other hand, the left hand can also be directly evaluated using the OPEs as 
\begin{align}\label{jjvvm2}
&\;\;\;\;\left\langle
\Bigl(j^+(z)-2x_j j^3(z)+x_j^2 j^-(z)\Bigr)
V^{w}_{j,h}(x_1; z_1) V^{w}_{j,h}(x_2; z_2) 
\right \rangle \\
&=  \sum_{i\neq j}\left( \frac{2 (x_i-x_j) h_i +(x_i-x_j)^2 \partial_{x_i}}{(z-z_i)} 
\left\langle V^{w}_{j,h}(x_1; z_1) V^{w}_{j,h}(x_2; z_2) \right \rangle
+   \sum_{\ell=1}^{w} \frac{(x_i-x_j)^2}{(z-z_i)^{\ell+1}} F^i_{\ell} \right) \nonumber \,.
\end{align}
\eqref{jjvvm1} means that all the singular terms in \eqref{jjvvm2} as $z\to z_i$ should vanish, and hence we obtain
\begin{equation}
\begin{aligned} 
\sum_{\ell=1}^{w} \frac{F^2_{\ell} }{(z-z_2)^{\ell+1}} 
&= -
 \frac{2 (x_2-x_1) h +(x_2-x_1)^2 \partial_{x_2}}{(z-z_2)(x_2-x_1)^2} 
\left\langle V^{w}_{j,h}(x_1; z_1) V^{w}_{j,h}(x_2; z_2) \right \rangle   \\
 &={1\over z-z_2}\left[
     \frac{2h}{(x_2-x_1)} +  \partial_{x_2}
       \right]
       \left\langle V^{w}_{j,h}(x_1; z_1) V^{w}_{j,h}(x_2; z_2) \right \rangle =0 
\end{aligned}
\end{equation}
where we have used the expression of the undeformed two-point function \eqref{x2pf} in the last step.
Similarly, we can show that 
\begin{equation} \label{eqn:Fs0}
    \sum_{\ell=1}^{w} \frac{F^2_{\ell} }{(z-z_2)^{\ell+1}}  = \sum_{\ell=1}^{w} \frac{F^1_{\ell} }{(z-z_1)^{\ell+1}}=\sum_{\ell=1}^{w} \frac{\bar{F^1}_{\ell} }{(\bar z-\bar z_1)^{\ell+1}}=\sum_{\ell=1}^{w} \frac{\bar{F^2}_{\ell} }{(\bar z-\bar z_2)^{\ell+1}}=0.
\end{equation}

\section{Single-trace $T\bar{T}$-deformed two-point functions}\label{sec:2pt-ttbar}
In this appendix we use the 
method of point-splitting to calculate the linearized perturbation to the two-point function of twist operatots. See \cite{Cardy:2019qao} and appendix D in \cite{Hirano:2020ppu} for this method.

Let us consider the first-order two-point functions of the twist operators under the single-trace $T\bar{T}$-deformation
\begin{equation}
    \begin{aligned}
    C_2=\frac{\mu}{\pi}\sum_{I=1}^n\int d^{2}x\langle T_{I}(x+\epsilon)\bar{T}_{I}(\bar{x})\sigma_{(12\cdots n)}(0)\sigma_{(n\cdots 21)}(a)\rangle,
    \end{aligned}
\end{equation}
where we have introduced the point-splitting parameter $\epsilon$. Since the stress-momentum tensor is conserved, it is useful to rewrite it by $T_I(x)=\partial_x\chi_I(x)$ and $ \bar{T}_I(\bar{x})=\partial_{\bar{x}}\bar{\chi}_I(\bar{x})$, which are correct except at the insertion point of the operator. The correlator $C_2$ thus can be written in terms of contour integral
\begin{equation}\label{c2chi}
    C_2=i\frac{\mu}{\pi}\sum_{I=1}^n\langle\int_{\partial S}d\bar{x}\Big(\chi_{I}(x+\epsilon)\bar{T}_{I}(\bar{x})\sigma_{(12\cdots n)}(0)\sigma_{(n\cdots 21)}(a)\Big)\rangle,
\end{equation}
where $\partial S$ is the contour encircling the possible singularity of $T(x)$ or $\chi(x)$. 
Using the the Ward identify \eqref{eq:Ward-CFT} and the explicit expression the Schwartzain \eqref{sch}, we can get the OPE around the two-points $x_1=0,\,x_2=a$ as,   
\be\label{opeTT}
 \sum_{I=1}^n  T_I(x)\bar{T}_I(\bar x)\sigma_{12\cdots n}(x_i)= \sum_{I=1}^n{c h_n\over 6 n}\Big({1\over 2(x-x_i)^2}\pm{1\over a (x-x_i)}+{O}((x-x_i)^0)\Big) \{\bar t_I(\bar x),\bar x\}\sigma_{12\cdots n}(x_i).
\ee
where the plus  sign is taken for $x_i=0$, and minus sign for $x_i=a.$
After integration, we 
get 
\be\label{chiTs}
\sum_{I=1}^n \chi_I(x)\bar{T}_I(\bar x)\sigma_{12\cdots n}(x_i)= \sum_{I=1}^n{c h_n\over 6 n}\Big(-{1\over 2(x-x_i)}\pm{1\over a }\ln(x-x_i)+{O}((x-x_i)^0)\Big) \{\bar t_I(\bar x),\bar x\}\sigma_{12\cdots n}(x_i).
\ee
from which we can explicitly see that $\chi_I(x+\epsilon)$ is not single valued around $0$ and $a$. As a result, the contour along $\partial S$ in \eqref{c2chi} can be divided into coutour integrals around the two branching points at $x_i=0,\,a$, and the contribution from the branch cut in between. Schematically, we have the following two types of contributions,  
\begin{align}\label{eq:sou-int}
\raisebox{-24.5pt}{%
\begin{tikzpicture}[scale=.75]
\draw[-<,thick] (1,0.1) -- (-0.05,0.1);
\draw[thick] (0,0.1) -- (-1,0.1,0);
\draw[-<,thick] (-1,-0.1) -- (0.05,-0.1);
\draw[thick] (0,-0.1) -- (1,-0.1);
\draw[-<,thick] (-1.58,-0.05) -- +(0,-0.01);
\draw[thick] (-1,0.1) arc (20:340:0.3);
\draw[thick] (1,-0.1) arc (-160:160:0.3);
\draw[fill=black] (-1.28,0) circle (0.1) node [left,xshift=-6] {$-\epsilon$};
\draw[fill=black] (1.28,0) circle (0.1) node [left,xshift=40] {$a-\epsilon$};
\node () at (0,-.9) {$\partial S$};
\end{tikzpicture}}
~~~~=~~~~
\raisebox{-25.5pt}{%
\begin{tikzpicture}[scale=.75]
\draw[-<,thick] (-1.58,-0.05) -- +(0,-0.01);
 \draw[thick] (-1,0.1) arc (20:340:0.3);
 \draw[fill=black] (-1.28,0) circle (0.1) node [left,xshift=-7] {$-\epsilon$};
 \node () at (-1.25,-.9) {(2)};
\end{tikzpicture}}
~~~~ + ~~~~
\raisebox{-25.5pt}{%
\begin{tikzpicture}[scale=.75]
\draw[-<,thick] (1,0.1) -- (-0.05,0.1);
\draw[thick] (0,0.1) -- (-1,0.1,0);
\draw[-<,thick] (-1,-0.1) -- (0.05,-0.1);
\draw[thick] (0,-0.1) -- (1,-0.1);
\node () at (0,-.9) {(1)};
\end{tikzpicture}}
~~~~+~~~~
\raisebox{-25.5pt}{%
\begin{tikzpicture}[scale=.75]
\draw[->,thick] (-1.,-0.05) -- +(0,-0.01);
\draw[thick] (-1.55,-0.1) arc (-160:160:0.3);
\draw[fill=black] (-1.28,0) circle (0.1) node [left,xshift=45] {$a-\epsilon$};
\node () at (-1.25,-.9) {(2)};
\end{tikzpicture}}.
\end{align}
Let us first consider the contribution from the branch cuts i.e. the first part $(1)$ in the above equation. Since the discontinuity is due to the insertion points at $x_i=0,\,a$, we  can evaluate it using the expansion \eqref{chiTs}, which amounts to
\begin{equation}
    \begin{aligned}
        C_{2,(1)}
&=i\frac{\mu}{\pi}\sum_{I,i}\int_{\bar{X}-\epsilon}^{\bar{x}_{i}-\epsilon}d\bar{x}\langle\chi(x+\epsilon)\bar{T}_{I}(\bar{x})\sigma_{(12\cdots n)}(0)\sigma_{(n\cdots 21)}(a)\rangle\big|^{{\rm below}}_{{\rm above}}\\
&=-\mu\frac{ch_{n}}{3na}\sum_{I=1}^{n}\int_{\bar{a}-\epsilon}^{\bar{0}-\epsilon}d\bar{x}\{\bar{t}_{I}(\bar{x}),\bar{x}\}\langle\sigma_{(12\cdots n)}(0)\sigma_{(n\cdots21)}(a)\rangle\\
   &= \frac{16\mu h_{n}^{2}}{n|a|^2}\log(\frac{\bar{a}}{\epsilon})\langle\sigma_{(12\cdots n)}\sigma_{(n\cdots21)}(a)\rangle +{O}(\epsilon).
   \label{c1f} \end{aligned}
\end{equation}
The second part $C_{2,(2)}$ is
\begin{equation}
\begin{aligned}
   C_{2,(2)}=&i\frac{\mu}{\pi}\sum_{I,i}\oint_{\bar{x}={\bar{x}}_{i}-\epsilon}d\bar{x}\langle\chi_{I}(x+\epsilon)\bar{T}_{I}(\bar{x})\sigma_{(12\cdots n)}(0)\sigma_{(n\cdots21)}(a)\rangle\\
   =&-\mu\sum_{I=1}^{n}\frac{4h_{n}}{n\bar{a}}(\int_{\bar{X}-\epsilon}^{0-\epsilon}dx-\int_{\bar{X}-\epsilon}^{{\bar{a}}-\epsilon}dx)T_I(x)\langle\sigma_{(12\cdots n)}(0)\sigma_{(n\cdots21)}(a)\rangle,
\end{aligned}
\end{equation}
where we have used the OPE between $\bar T$ and the twist operators, carried out the contour integrals enclosing $0$ and $a$, and rewritten $\chi_I(x)$ in terms of the integration of $T(x)$. The two integrals in the above equation combines, which using the Schwartzian again, leads to the final contribution from the two branching points
\begin{equation}
    \begin{aligned}
       C_{2,(2)}=&-\mu\frac{ch_{n}}{3n\bar{a}}\sum_{I=1}^{n}\int_{a-\epsilon}^{0-\epsilon}\{t_{I}(x),x\}\langle\sigma_{(12\cdots n)}(0)\sigma_{(n\cdots21)}(a)\rangle\\
       =&\frac{16\mu h_{n}^{2}}{n|a|^{2}}\log(\frac{a}{\epsilon})\langle\sigma_{(12\cdots n)}(0)\sigma_{(n\cdots21)}(a)\rangle+{O}(\epsilon).\label{c2f}
    \end{aligned}
\end{equation}
Putting together the two parts \eqref{c1f} and \eqref{c2f}, we get the first-order single-trace $T\bar{T}$ deformed two-point function 
\begin{equation}
    C_2=\frac{ 16\mu h_{n}^{2}}{n|a|^2} \log(\frac{|a|^{2}}{\epsilon^{2}})\langle\sigma_{(12\cdots n)}(0)\sigma_{(n\cdots21)}(a)\rangle+{ O}(\epsilon),
\end{equation}
which reproduces the result  \eqref{eq:first order-renom} obtained from dimensional regularization and renormalizarion.

\bibliographystyle{ssg}
\bibliography{BIBLIOGRAPHY}

\end{document}